\documentclass[floatfix,showpacs,pra,aps,twocolumn]{revtex4}

\usepackage{bm}
\usepackage{amsmath}
\usepackage{amssymb}
\usepackage{latexsym}
\usepackage{amsfonts}
\usepackage{epsfig}
\usepackage{psfrag}

\newcommand{\ket}[1]{\left|#1\right>}
\newcommand{\bra}[1]{\left<#1\right|}
\newcommand{\braket}[2]{\left<#1|#2\right>}
\newcommand{\expval}[1]{\left< #1 \right>}
\newcommand{\nn}{\nonumber\\}

\newcommand{\f}[1]{\mbox{\boldmath$#1$}}

\newcommand{\bea}{\begin{eqnarray}}
\newcommand{\ea}{\end{eqnarray}}
\newcommand{\eea}{\end{eqnarray}}
\newcommand{\ord}{{\cal O}}
\newcommand{\abs}[1]{{\left| #1 \right|}}

\begin{document}

\title{The role of symmetries in adiabatic quantum algorithms}

\author{Gernot Schaller$^1$ and Ralf Sch\"utzhold$^2$}

\affiliation{1 Institut f\"ur Theoretische Physik, Hardenbergstra{\ss}e 36,
Technische Universit\"at Berlin, D-10623 Berlin, Germany\\
2 Institut f\"ur Theoretische Physik, Lotharstra{\ss}e 1,
Universit\"at Duisburg-Essen, D-47048 Duisburg, Germany}

\begin{abstract}
Exploiting the similarity between adiabatic quantum algorithms and
quantum phase transitions, we argue that second-order transitions --
typically associated with broken or restored symmetries -- should be
advantageous in comparison to first-order transitions.
Guided by simple examples we construct an alternative adiabatic algorithm
for the NP-complete problem {\em Exact Cover 3}.
We show numerically that its average performance (for the considered cases up to $\ord\{20\}$
qubits) is better than that of the conventional scheme. 
The run-time of adiabatic algorithms is not just determined by the minimum 
value of the fundamental energy gap (between the ground state and the
exited states), but also by its curvature at the critical point.  
The proposed symmetry-restoring adiabatic quantum algorithm only
contains contributions linear and quadratic in the Pauli matrices and
can be generalized to other problem Hamiltonians which are
decomposed of terms involving one and two qubits.
We show how the factoring problem can be cast into such a quadratic
form.  
These findings suggest that adiabatic quantum algorithms can solve a
large class of NP problems much faster than the Grover search routine 
(which corresponds to a first-order transition and yields a quadratic
enhancement only).
\end{abstract}


\maketitle

\section{Introduction}

\subsection{Adiabatic quantum algorithms}

The overwhelming potential of quantum computers in number
factorization \cite{shor} and database search \cite{grover} has
initiated a lot of research.
In the conventional (sequential) quantum algorithmic approach, unitary 
single and two-qubit operations are applied on a quantum state and the
result of the computation can then be extracted (probabilistically) by 
measurement. 
Unfortunately, the promises of quantum computing are strongly hampered
by the devastating influence of decoherence: 
Inevitable couplings with the reservoir tend to destroy the fragile
quantum coherence.
The available error correction schemes require further 
auxiliary qubits \cite{nielsen2000} and -- although even strongly 
correlated errors can in principle be corrected \cite{aliferis2006a} -- become 
increasingly complex with admitting more and more general errors.

Farhi {\em et al.} have proposed an alternative scheme with 
an inherent error protection: adiabatic quantum computation
\cite{farhi,farhi_ax}. 
In this scheme, the solution to a problem is encoded as the (unknown)
ground state of a (known) problem Hamiltonian $H_{\rm F}$, which is
separated by a sufficiently large energy gap from excited states.
Computation will start with a different initial Hamiltonian 
$H_{\rm I}$ possessing a known and easily preparable ground state, 
which is also well separated energetically from all other states.
The system is prepared in this ground state and then the initial
Hamiltonian $H_{\rm I}$ is continuously deformed into the final
problem Hamiltonian $H_{\rm F}$.  
The adiabatic theorem guarantees that if the evolution is slow enough, the system will stay
near its instantaneous eigenstate and will thus finally reach the
ground state of the problem Hamiltonian \cite{sarandy2004}. 
Thus, the solution to the problem can be extracted by measurement of
the final quantum state.
A nice advantage of this quantum computation scheme (which is believed to
be polynomially equivalent to sequential quantum computation
\cite{aharonov}) lies in the robustness of the ground state
against the influences of decoherence 
-- a sufficiently cold reservoir provided 
\cite{childs2001a,sarandy2005a,sarandy2005b,roland2005a,aberg2005a,aberg2005b,thunstroem2005a,shenvi2003a,tiersch,mostame2007a}:
The ground state cannot decay and phase errors do not play any role,
i.e., errors can only result from excitations. 
The speed (total evolution time $T$) at which the interpolation
between the two Hamiltonians can be performed without strongly
exciting higher eigenstates of the system is related to the inverse
gap between the instantaneous lowest and first excited eigenvalues
\cite{sarandy2004}.  
The total evolution time $T$ required to reach a fixed fidelity in the
final state can be interpreted as the algorithmic complexity of the
adiabatic computation scheme. 
With using a constant deformation speed (ignoring the structure of the
fundamental energy gap) the adiabatic run-time will scale as
$T=\ord\{g_{\rm min}^{-2}\}$, whereas with knowledge on the
fundamental gap this can under fairly moderate assumptions even be improved to
$T=\ord\{g_{\rm min}^{-1}\}$ \cite{schaller2006b,jansen2007a}.

This also highlights the main obstacle in adiabatic quantum
computation: 
Typically, the minimum fundamental gap $g_{\rm min}$ decreases
strongly with increasing system size (number of qubits) for nontrivial
problems.  
For example, in the adiabatic version of Grover's algorithm
\cite{roland2002}, the minimum fundamental gap decreases exponentially 
$g_{\rm min} = 2^{-n/2}$ with system size $n$.
An analogous exponential scaling of the minimum fundamental gap has
been found in other adiabatic quantum algorithms with similar initial
Hamiltonians \cite{znidaric+horvat}.
However, it has been argued that this exponential scaling is a result
of the unfavorable choice of the initial Hamiltonian and can be
avoided for more suitable choices \cite{farhi-fail}. 
In these cases, however, $g_{\rm min}$ is not known analytically.
Numerical analysis in \cite{farhi,farhi_ax} seems to indicate a quadratic
scaling  $T \sim n^2$ of the algorithmic complexity with the
system size $n$ -- but this favorable scaling could perhaps be just a 
consequence of choosing particularly simple problems,
cf.~\cite{znidaric,mosca}.   
Therefore, the speed-up attainable with these adiabatic quantum 
algorithms is still an open question. 

\subsection{Quantum phase transitions}

As has been noted earlier \cite{latorre2004a,quantum_phase}, adiabatic  
quantum algorithms display a remarkable similarity with sweeps through
quantum phase transitions: 
During the adiabatic interpolation, the ground state changes from the
simple initial ground state of $H_{\rm I}$ to the unknown solution
of some problem encoded in $H_{\rm F}$.  
Typically, on the way from $H_{\rm I}$ to $H_{\rm F}$, one encounters
a critical point where the fundamental gap (which is
sufficiently large initially and finally) becomes very small.  
Near the position of the minimum gap, the ground state will change 
more drastically than during other time intervals of the interpolation
\bea\label{Egs_general}
||\dot\psi_0||^2\equiv
\braket{\dot\psi_0}{\dot\psi_0}\geq
\sum\limits_{n>0}
\left|\frac{\bra{\psi_0}\dot H\ket{\psi_n}}{E_n-E_0}\right|^2
\,,
\ea
where we have inserted an identity and expressed $\braket{\dot\psi_0}{\psi_n}$ by the time-derivative of the eigenvalue equation 
$H(t)\ket{\psi_0(t)}=E_0(t)\ket{\psi_0(t)}$.
If the relevant fundamental energy gap (the smallest one with 
\mbox{$\bra{\psi_0}\dot H\ket{\psi_n} \neq 0$}) 
scales inversely with the system size,
we see directly that in the infinite-size limit, the ground state will change 
non-analytically at the critical point.
This singularity would completely prohibit an adiabatic evolution in this limit.
However, for practical problems one is interested in the finite-size
scaling of an algorithm, where it makes a huge difference whether its
computational complexity increases {\em exponentially} or merely 
{\em polynomially} with the system size $n$.
The same principle applies to the efficient suppression of thermal excitations when
the quantum system is coupled to a low-temperature bath \cite{childs2001a}, where it may prove
experimentally difficult to apply polynomially small temperatures and most likely infeasible to
apply exponentially small temperatures.

In the following, we shall exploit the analogy between adiabatic 
quantum algorithms and quantum phase transitions further in order to
gain additional insight into these issues. 
Beyond the behaviour of the minimum gap, the complexity of
implementing the involved Hamiltonians (number of interactions between
different qubits) is of experimental importance and shall also be
addressed.

\section{Motivation}\label{Smotivation}

\subsection{Grover Search Routine}\label{SSgrover}

The adiabatic version of Grover's search algorithm is defined by the 
linear interpolation 
\mbox{$H(s) = \left(1-s\right) H_{\rm I} + s H_{\rm F}$}, 
where $s \in [0,1]$ between the Hamiltonians \cite{roland2002} 
\bea
\label{Ehamgrover}
H_{\rm I} &=& \f{1}-\ket{S}\bra{S}\,,\qquad
H_{\rm F} = \f{1}-\ket{w}\bra{w}\,,
\eea
where the initial ground state
\bea
\label{Etotalsup}
\ket{S} = \frac{1}{\sqrt{N}} \sum_{z=0}^{N-1} \ket{z} =
\ket{\to}\otimes\ldots\otimes\ket{\to} 
\eea
is the superposition vector of all $N=2^n$ states in 
the computational ($\sigma_z$) basis and 
\mbox{$\ket{\to} = (\ket{0}+\ket{1})/\sqrt{2}
= (\ket{\downarrow}+\ket{\uparrow})/\sqrt{2}$}.
In contrast, the final ground state $\ket{w}$ is some distinguished
computational basis state.

\begin{figure}[ht]
\includegraphics[height=4cm,clip=True]{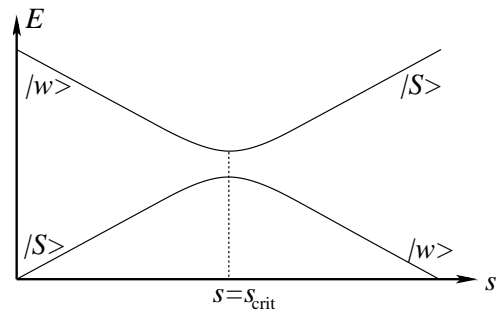}
\caption{\label{Ffirst_spectrum} Sketch of the two lowest energy
  levels for the Grover problem as a prototypical example for a
  first-order quantum phase transition in the infinite size limit
  $n\uparrow\infty$ (where the energy gap vanishes and the avoided
  level crossing becomes a real crossing).}
\end{figure}

It is straightforward to derive the instantaneous spectrum of $H(s)$ 
by Erhard-Schmidt orthogonalization, for example. 
One obtains the fundamental gap 
\bea
\label{Especgrover}
g(s) = \sqrt{1-4\left(1-\frac{1}{2^n}\right)s(1-s)}\,,
\eea
and the two non-trivial levels 
\mbox{$E_{0/1}(s) = \frac{1}{2}\left[1\pm g(s)\right]$}
as well as 
\mbox{$E_2(s) = \ldots = E_{N-1}(s)=1$}.
Thus, the adiabatic Grover algorithm continuously transforms the
initial vacuum $\ket{S}$ towards the final ground state $\ket{w}$, see
also figure \ref{Ffirst_spectrum}. 
From Equation (\ref{Especgrover}) we may infer an exponentially small
minimum fundamental gap $g_{\rm min}=\sqrt{1/2^n}$ at the critical
point $s_{\rm crit}=1/2$. 

In the infinite-size limit ($n\to\infty$), the first derivative 
$dE_0/ds$ of the ground state energy $E_0(s)$ jumps
at the point of the phase transition $s_{\rm crit}=1/2$. 
In view of the equality 
\bea
\label{jump}
\frac{dE_0(s)}{ds}=
\left<\psi_0(s)\left|\frac{d{H}(s)}{ds}\right|\psi_0(s)\right>
\,,
\ea
this discontinuity goes along with a jump in certain expectation
values (e.g., $\expval{dH/ds}$) and hence order parameters such as the
horizontal magnetization $\langle\sigma_x\rangle$ change abruptly at
the critical point $s_{\rm crit}=1/2$.  
Conventionally, this property (discontinuous order parameter) is used 
to classify the phase transition as a first-order transition.

\begin{figure}[ht]
\begin{tabular}{c}
\includegraphics[height=6cm,clip=True]{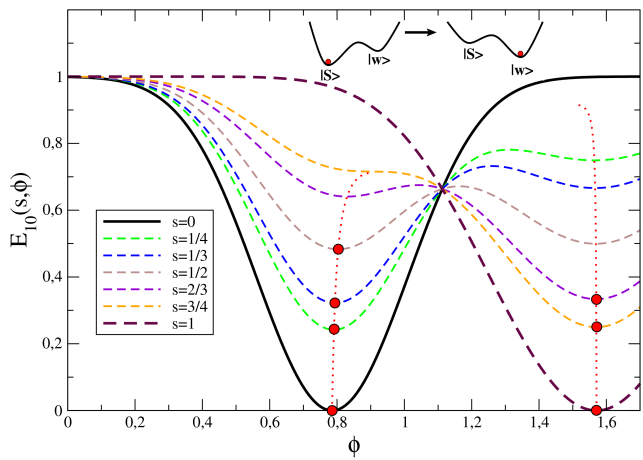}
\end{tabular}
\caption{\label{Fsketch_grover} 
Semiclassical energy landscapes (\ref{Esketch_grover}) 
for the Grover problem with $n=10$ spins for different interpolation parameters $s$. 
In order to stay in the global minimum, the system has to tunnel through an
energy barrier that does not vanish throughout the interpolation, i.e., the time-dependent 
position (red dotted lines) of the two local minima is not connected. Circle symbols show the
position of the global minimum.}
\end{figure}

From Eqn. (\ref{Egs_general}) it follows that the ground state basically 
jumps from 
\mbox{$\ket{\psi_0(s<1/2)}=\ket{S}$} to  
\mbox{$\ket{\psi_0(s>1/2)}=\ket{w}$}. 
Such an abrupt change is a general feature of first-order quantum phase transitions
and can be understood in terms of a time-dependent energy landscape.
Assuming a separable state (compare also \cite{farhi0201031,farhi0208135} for a similar approach)
\bea\label{Eseparable}
\ket{\varphi}=\bigotimes\limits_{i=1}^n \left[\cos(\varphi)\ket{0}+\sin(\varphi)\ket{1}\right]_i
\eea
and re-ordering the states in the computational basis such that $\ket{w}=\ket{1\ldots 1}$
(the initial state (\ref{Etotalsup}) is invariant to this transformation)
we can calculate the semiclassical energy landscape 
\bea\label{Esketch_grover}
E(s,\varphi) &\equiv& \bra{\varphi} H(s) \ket{\varphi}\nn
&=& 1 - (1-s) \frac{\left[\cos(\varphi)+\sin(\varphi)\right]^{2n}}{2^n}\nn
&& - s \sin^n(\varphi)\,,
\eea
shown in Fig.~\ref{Fsketch_grover}.
%
%
We see that even in this semiclassical picture the two minima are always separated -- either by 
a potential barrier or by large energy differences.
This corresponds to the existence of a tunneling barrier between the
two minima.

The time-dependent Hamiltonian (\ref{Ehamgrover}) does not have a conserved quantity, such
that the initial ground state $\ket{S}$ does not break any symmetry of the initial Hamiltonian.
We conjecture that in the absence of additional conserved quantities 
(related to continuous symmetries of the Hamiltonian), there are always 
two (or more) competing local minima whose energies depend on the interpolation
parameter $s$. 
At the critical point $s_{\rm crit}$, the order of the $s$-dependent
energy minima changes and thus the global minimum makes a jump.
Consequently, in order to stay in the true ground state, the quantum
system has to tunnel through the barrier depicted in figure
\ref{Fsketch_grover}. 
Since one would naturally expect that all the energy scales and hence
also the size of the tunneling barrier increase roughly linearly
with system size $n$, the tunneling probability decreases
exponentially with the number of qubits $n$. 
This intuitive picture suggests that the run-time $T$ needed to stay
in the ground state scales exponentially for all first-order
transitions.
Indeed, since the tunneling amplitude provides the coupling between
the competing ground states and hence determines the minimum gap
$g_{\rm min}$ in their avoided level crossing, such an exponential
scaling is precisely what one finds for the adiabatic Grover
algorithm and further adiabatic algorithms 
\cite{znidaric+horvat,mosca,farhi-fail} with similar initial
Hamiltonian -- which all correspond to first-order transitions. 

\subsection{Transverse Ising model}\label{SSising}

\begin{figure}[ht]
\begin{tabular}{c}
\includegraphics[height=4cm,clip=True]{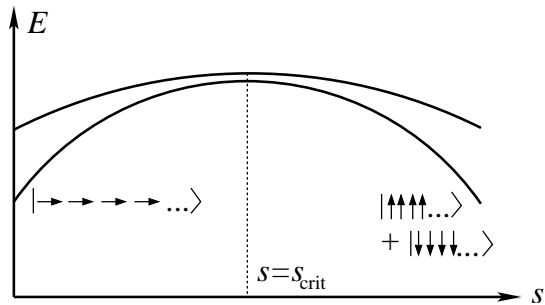}
\end{tabular}
\caption{\label{Fsecond_spectrum}
Sketch of the two lowest energy eigenvalues for the Ising model as a 
prototypical example of a second-order quantum phase transition. In contrast to figure
\ref{Ffirst_spectrum}, the geometry near the minimum gap does not correspond to an
isolated avoided crossing.}
\end{figure}

After having discussed first-order transitions, let us turn to 
a prototypical model for a second-order quantum phase
transition \cite{sachdev}. 
The one-dimensional transverse Ising model is given by linear 
interpolation between 
\bea
\label{Ehamising}
H_{\rm I} &=& -\sum_{\ell=1}^n \sigma^x_\ell\,,\qquad
H_{\rm F} = -\sum_{\ell=1}^n \sigma^z_\ell\sigma^z_{\ell+1}\,,
\eea
where $\sigma^{x/y}_\ell$ denote the Pauli spin matrices acting on the
$\ell^{\rm th}$ qubit and periodic boundary conditions
$\sigma^z_{n+1}=\sigma^z_1$ are assumed. 
At any point during the interpolation, the Hamiltonian $H(s)$ can be
diagonalized analytically: 
The successive application of Jordan-Wigner, Fourier, and Bogoliubov 
transformations \cite{sachdev} map the interacting
spin-1/2-Hamiltonian (\ref{Ehamising}) to a set of non-interacting
fermionic quasi-particles  
\bea
H = 
\sum_k \epsilon_k \left(\gamma_k^\dagger \gamma_k -
\frac{1}{2}\right)\,, 
\eea
with fermionic creation and annihilation operators 
$\gamma_k^\dagger,\gamma_k$ and single quasi-particle energies 
\bea
\label{Esinglepart}
\epsilon_k(s) = 2\sqrt{1-4 \cos^2(ka/2)s(1-s)}\,,
\eea
see figure \ref{Fsecond_spectrum}.
%
The wavenumber $ka$ covers the range 
\mbox{$ka=\left(1+2\mathbb{Z}\right)\pi/n 
\quad : \quad \abs{ka}<\pi$}.
Both Hamiltonians in (\ref{Ehamising}) obey a 180-degree rotational
symmetry around the $\sigma^x_\ell$-axes (bit-flip)
\bea
\label{Ebit-flip}
\left[H_{\rm I},\bigotimes\limits_{\ell=1}^n \sigma^x_\ell\right]=
\left[H_{\rm F},\bigotimes\limits_{\ell=1}^n \sigma^x_\ell\right]=0
\ea
which transforms $\sigma^z_i$ to $-\sigma^z_i$. 
Since due to Eqn. (\ref{Ebit-flip}) the even bit-flip parity of the
unique initial ground state $\ket{S}$ is a conserved quantity, 
the symmetry of the two-fold degenerate final ground state 
$\ket{\Psi_0^{1}} = \ket{\uparrow\ldots\uparrow}$ 
and 
$\ket{\Psi_0^{2}}=\ket{\downarrow\ldots\downarrow}$
is broken, such that for adiabatic evolution, the system will end
up in the 
macroscopic superpositions (Schr\"odingers cat) state 
\bea
\label{cat}
\ket{\Psi_0^{\rm even}}=
\frac{\ket{\uparrow\ldots\uparrow}+\ket{\downarrow\ldots\downarrow}}
{\sqrt{2}} 
\,.
\ea
The minimum gap can be obtained from equation (\ref{Esinglepart}) and
scales polynomially \mbox{$g_{\rm min}=\ord\{1/n\}$}.
Therefore, the  adiabatic runtime $T$ does also scale polynomially --
a constant speed interpolation, for example, yields  
$T_{\rm ad}=\ord\{n^2\}$ \cite{dziarmaga}. 
Furthermore, since the wavenumber $ka$ covers the range 
\mbox{$ka=\left(1+2\mathbb{Z}\right)\pi/n\quad : \quad\abs{ka}<\pi$}, 
the infinite-size limit of the ground state energy can be obtained by 
replacing the sum over the single-particle energies by an integral 
\bea
E_0^{\rm cont}(s) 
&=& 
-\lim_{n\to\infty} \sum_{ka} \frac{\epsilon_k}{2}
= -\frac{2n}{\pi} {\cal E}\left(2 \sqrt{s(1-s)}\right)
\,,\qquad
\eea
where ${\cal E}(x)$ denotes the complete elliptic integral.
In contrast to the previous subsection \ref{SSgrover}, the energy density  
${\cal E}(1)=1$ and its first derivative ${\cal E}'(1)=0$ are well
defined at the critical point. 
Since $\langle H(s)\rangle$ and $\langle dH/ds\rangle$ are continuous
across the transition, there is no jump in order parameters such as 
$\langle\sigma_x\rangle$ and the ground state changes less abruptly. 
However, the second derivative of the ground state energy diverges at
the critical point  ${\cal E}''(1)=\infty$ and hence,
we classify the Ising model as a second order quantum phase
transition, compare also \cite{sachdev}. 
Consequently, we have a symmetry-breaking quantum phase transition of
second order.

\begin{figure}[ht]
\begin{tabular}{c}
\includegraphics[height=6cm,clip=True]{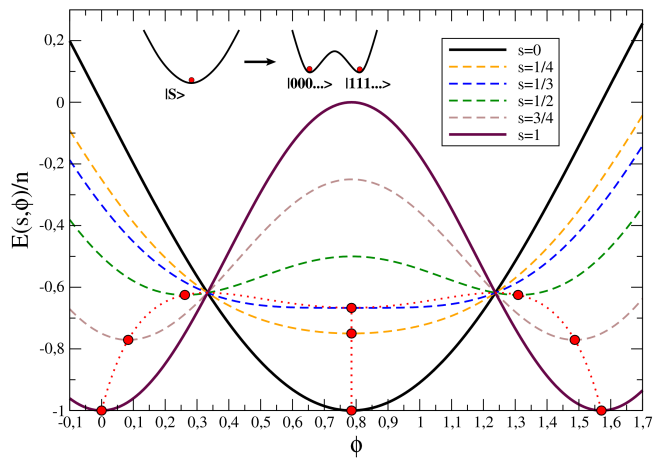}
\end{tabular}
\caption{\label{Fsketch_ising}
Semiclassical energy density landscape for the Ising model in a transverse field 
for different interpolation parameters $s$.
In contrast to first-order transitions, there is no energy barrier
between the initial and the final ground state, i.e., the time-dependent positions
(red dotted line) of the minima are always connected.
}
\end{figure}

As in the previous section, we can derive a semiclassical time-dependent
energy landscape with the separable ansatz (\ref{Eseparable}) to obtain
\bea\label{Esketch_ising}
E_n(s,\varphi) &=& n\Big\{
-(1-s) \sin(2\varphi)\nn
&&- s \left[ \cos^2(\varphi) - \sin^2(\varphi)\right]^2\Big\}\,,
\eea
where the  bit-flip invariance (\ref{Ebit-flip}) is reflected by a mirror symmetry of the
energy landscape around $\varphi=\pi/4$, see figure \ref{Fsketch_ising}.
%
%
We see that symmetry-breaking/restoring quantum phase transitions are typically of
second (or higher) order, i.e., the ground state does not change
as abruptly as in first-order transitions. 
Consequently, the system does not need to tunnel through a barrier in
order to stay in the global vacuum. 
As a result, the reason for the exponential scaling of the run-time
$T$ we found for first-order transitions is absent in the case of quantum
phase transitions of second or higher order, which is consistent with the
polynomial scaling of gap \cite{sachdev} and adiabatic runtime \cite{dziarmaga}.

Evidently, the transverse Ising model can also be considered an
adiabatic quantum algorithm, although of course the final ground
states are trivial -- the problem to be solved is simply 
``How can all bits have the same value?''
Note that, reversing the evolution and slowly interpolating from
$H_{\rm F}$ towards $H_{\rm I}$ in (\ref{Ehamising}), the final state
will  be close to the true unique ground state $\ket{S}$ if and only
if initialized with the superposition (\ref{cat}).

\subsection{Mixed Case}\label{SSMixed}

\begin{figure}[ht]
\begin{tabular}{c}
\includegraphics[height=6cm,clip=True]{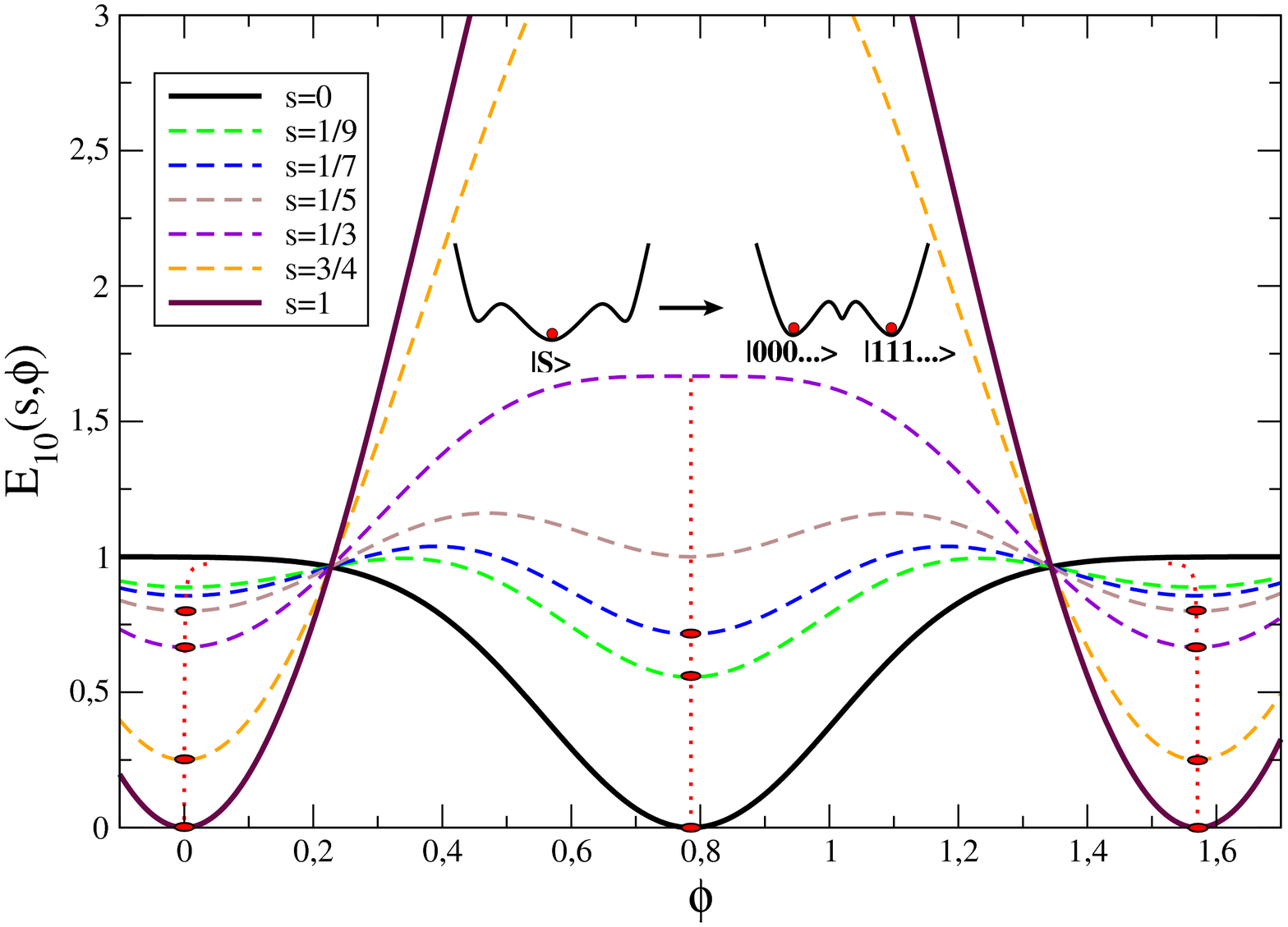}
\end{tabular}
\caption{\label{Fsketch_mixed}
Semiclassical energy landscapes for a symmetry-breaking quantum phase transition -- 
which is, however, not of second but of first order. In spite of the symmetry breaking, there
is a tunneling barrier throughout the interpolation
and a jump between the initial and the final
ground state(s), i.e., the position of the vacua (dotted red lines) is not connected.}
\end{figure}

\begin{figure}[ht]
\begin{tabular}{c}
\includegraphics[height=6cm,clip=True]{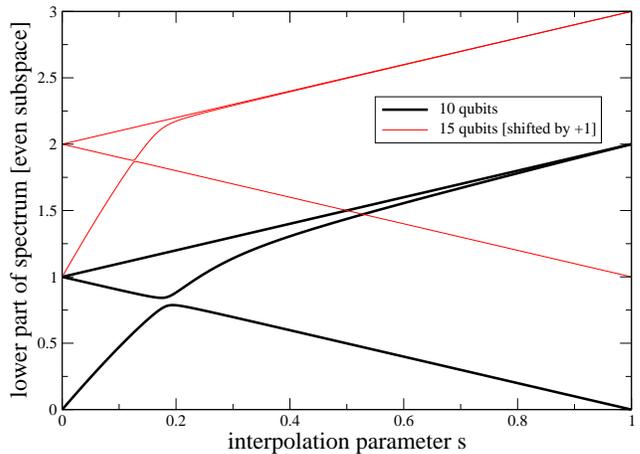}
\end{tabular}
\caption{\label{Fthird_spectrum} Lowest eigenvalues of
  the Hamiltonian (\ref{H-mixed}) for 10 (bold black lines) and 15 (thin red
  lines, shifted for better readability) qubits in the relevant 
  subspace which is even under bit-flip. One can clearly see that the
  spectrum displays an avoided level crossing at the critical point --
  thus corresponding to a first-order transition.}
\end{figure}

Even though the concept of broken or restored symmetries is very
useful for classifying quantum phase transitions, it should be
stressed that symmetry breaking/restoration alone does not guarantee a
second-order transition. 
As an intuitive counterexample, one may consider the linear interpolation
between
\bea
\label{H-mixed}
H_{\rm I} &=& 
\f{1}-\ket{S}\bra{S}\,,\qquad
H_{\rm F} = 
\sum_{\ell=1}^n \frac{1}{2}\left[\f{1}-\sigma^z_\ell\sigma^z_{\ell+1}\right]
\,,\qquad
\eea
which has an equivalent (up to shifting and scaling to obtain positive definiteness) 
final Hamiltonian as (\ref{Ehamising}) but differs in the initial Hamiltonian, for 
which we have chosen the initial Hamiltonian of the Grover problem (\ref{Ehamgrover}).
Evidently, the ground state symmetry of the final Hamiltonian is also broken, since the
bit-flip parity is also conserved.
However, an analysis of the semiclassical energy landscapes generated with a 
separable ansatz (\ref{Eseparable}) yields
\bea
E_n(s, \varphi) &=& (1-s)\left\{1-\frac{\left[\cos(\varphi)+\sin(\varphi)\right]^{2n}}{2^n}\right\}\nn
&&+ \frac{n s}{2} \left\{1 - \left[ \cos^2(\varphi) - \sin^2(\varphi)\right]^2\right\}\,,
\eea
which also exhibits a mirror symmetry at $\varphi=\pi/4$ but always has a tunneling barrier between
the vacua, see figure \ref{Fsketch_mixed}.
%
The level structure displays the geometry of an
avoided level crossing at the critical point, i.e., it corresponds to
a first-order phase transition, see Figure \ref{Fthird_spectrum}.
In accordance with our previous arguments, it can be shown
analytically that the gap scales exponentially in such a situation 
\cite{znidaric+horvat,mosca,farhi-fail}. 

In order to understand why the phase transition is of first order, it
might be useful to recall that the initial Hamiltonian is a projector
that involves $n$-qubit interactions 
\bea
\label{n-qubit}
H_{\rm I} = \f{1}-\ket{S}\bra{S}
=\f{1}-\bigotimes_{\ell=1}^n\frac{\f{1}+\sigma_\ell^x}{2}
\,.
\eea
The bit structures of the initial and the final Hamiltonian are very
different and, in this sense, their ``distance'' in the space of all
matrices is very large. 

\subsection{Conjecture}\label{SSconjecture}

We have seen for the Grover example that first order quantum phase transitions 
are associated with an exponential scaling of the adiabatic runtime with the system size $n$.
For local Hamiltonians, this observation can be generalized \cite{schuetzhold2008a}: 
For a typical first order transition one has two locally distinguishable competing 
ground states $\ket{\tilde w}$ and $\ket{\tilde s}$ that exchange their energetic 
favorability right at the critical point 
$\left.\bra{\tilde w} H \ket{\tilde w}\right|_{s_{\rm crit}}=\left.\bra{\tilde s} H \ket{\tilde s}\right|_{s_{\rm crit}}$.
Their overlap is exponentially small (due to local distinguishability) and for a 
local reasonable (with polynomial matrix elements) Hamiltonian this also implies that 
$\bra{\tilde s} H \ket{\tilde w} \propto \exp\{-\ord(n)\}$ is also exponentially small.
In the two-dimensional subspace spanned by these states the energy gap equates to
$g_{\rm crit}= 2 \abs{\bra{s} H \ket{w}}$,
which becomes therefore exponentially small.

Our conjecture is therefore that for adiabatic quantum algorithms\\
a.) higher order quantum phase transitions are more advantageous 
than first order quantum phase transitions and\\
b.) such higher order transitions might be induced by interpolations 
respecting a conserved quantity (leading to spontaneous symmetry breaking
in the degenerate ground state of one Hamiltonian) in combination with
similar bit structures in initial and final Hamiltonians.

Of course, these considerations are very intuitive and by no means
conclusive -- but they will hopefully help us to design more powerful
quantum algorithms. 
For example, it would of course be nice to have no phase transition at all
(which corresponds to a constant lower bound on the fundamental energy gap \cite{schaller2008b})
but numerous numerical \cite{latorre2004a,banuls2006a,znidaric,quantum_phase,young2008a} 
and analytical \cite{znidaric+horvat,mosca} evidence suggests that this is not the case for
hard optimization problems.

\section{Runtime Scaling for Linear Quenching}\label{Sruntime}

The main measure for the computational complexity
of adiabatic quantum algorithms is the minimum runtime $T$. 
In the following, we derive an estimate of the runtime $T$
based on the level structure of the adiabatic quantum algorithm. 
We will assume a linear quench $s=t/T$ and a single isolated position of 
the minimum energy gap.
In cases where the position and size of this minimum energy gap is explicitly 
known, one can improve the runtime by adapting the interpolation speed 
(i.e., moving fast away from the critical point and slow in its
vicinity) \cite{roland2002,schaller2006b}.  
However, since the exact position of the minimum gap is not known 
{\em a priori} for most of the interesting cases -- although its
approximate position can be estimated by perturbative and other (such as 
e.g., Gershgorin's circle theorem) methods \cite{aharonov} --  
we shall work with a constant-speed interpolation.

As derived in \cite{schaller2006b}, the final occupation amplitude of
the first excited state $a_1(s=1)$ can be obtained formally from the
adiabatic expansion {\em via} 
\bea
\label{Eamplitude}
a_1(1) e^{-i \gamma_1(1)} &=& - \int\limits_0^1 ds\,a_0(s) e^{-i \gamma_1(s)} 
\frac{F_{01}(s)}{g(s)} \times\nn
&&\times\exp\left\{
-i T \int\limits_0^s g(s') ds'\right\}
\,,
\eea
where $F_{01}(s)$ denotes the transition matrix element of $H'(s)$ in  
the instantaneous energy basis, $a_{0}(s)$ the ground-state
amplitude, and $\gamma_1$ is a pure phase 
(including the Berry phase).  
The $ds$-integration along the real axis in the above integral can be
deformed in the lower complex half-plane \cite{schaller2006b}, 
where it is visible that for slow interpolations (large $T$), the outer 
integrand in (\ref{Eamplitude}) is exponentially suppressed (adiabatic approximation). 
However, such a deformation will be limited by singularities of
$g^{-1}(s)$ at $\tilde{s}$ located in the complex plane near the
minimum of $g(s)$ on the real axis. 
This determines how the adiabatic runtime $T$ has to scale with the
spectral properties in order to suppress the excitation amplitude in
(\ref{Eamplitude}) efficiently 
\bea\label{Econdition}
\Re
\left[
i T \int\limits_0^{\Re(\tilde{s})+i\Im(\tilde{s})/2} g(s) ds
\right] 
\gg 1
\,,
\eea
for a detailed discussion see \cite{schaller2006b}.
Assuming a sufficiently smooth behavior, we Taylor-Laurent expand the 
fundamental gap near its minimum (which is a saddle-point in the
complex plane) 
\bea
g(s) \approx g_{\rm min} + c_{\rm min} \left(s - s_{\rm crit}\right)^2\,,
\eea
where $g_{\rm min}$ denotes the value and $c_{\rm min}$ the curvature
of the fundamental gap at the critical point $s=s_{\rm crit}$.
The singularities of $1/g(s)$ are then found at approximately 
$\tilde{s}\approx s_{\rm crit} \pm i \sqrt{g_{\rm min}/c_{\rm min}}$,
which yields for the left hand side of (\ref{Econdition})
\bea
\Re\left[
i T 
\int\limits_0^{s_{\rm crit} - i \sqrt{g_{\rm min}/c_{\rm min}}/2}
g(s) ds
\right]
=\ord\left\{T \frac{g_{\rm min}^{3/2}}{c_{\rm min}^{1/2}}\right\}
\,.
\eea
Consequently, the runtime necessary to suppress the excitation amplitude
(\ref{Eamplitude}) efficiently scales as
\bea
\label{Escaling}
T = \ord\left\{\sqrt{\frac{c_{\rm min}}{g_{\rm min}^3}}\right\}
\,.
\eea
Evidently, the runtime does not only depend on the value of the
minimum gap $g_{\rm min}$ but also on its curvature $c_{\rm min}$ 
at the critical point $s=s_{\rm crit}$.
At first sight, this may seem counterintuitive, since from a naive 
interpretation of the adiabatic theorem (focusing on the minimum
gap only), one would expect that a small curvature  
(which implies a longer persistence of a small gap) 
should lead to longer run-times. 
However, it should be kept in mind that a large curvature means a
rapid change of the spectral characteristics of the Hamiltonian and
hence the system will find it harder to follow the evolution in order
to stay in the ground state \cite{schaller2006b,jansen2007a}. 

As a consistency check, we show that (\ref{Escaling}) correctly
reproduces the scaling of the adiabatic runtime found earlier  
for the Grover problem and the Ising model in case of
constant-speed interpolation: 
For the Grover model, one obtains from equation (\ref{Especgrover})
\bea
T \sim
\sqrt{\frac{g_{\rm G}''(s_{\rm crit})}{g_{\rm G}^3(s_{\rm crit})}}=
2N-1+\ord\left\{\frac{1}{N}\right\}\,,
\eea
which implies an exponential scaling of the adiabatic runtime 
(since $N=2^n$) with the system size, compare also
\cite{roland2002,schaller2006b} and subsection \ref{SSgrover}. 
Likewise, one obtains for the Ising model from equation
(\ref{Esinglepart})
\bea
T \sim
\sqrt{\frac{g_{\rm I}''(s_{\rm crit})}{g_{\rm I}^3(s_{\rm crit})}}= 
\frac{2}{\pi^2} n^2 -\frac{1}{12} +\ord\left\{\frac{1}{n}\right\}
\eea
a quadratic scaling of the adiabatic runtime, see also
\cite{dziarmaga} and subsection \ref{SSising}.

\begin{figure}[ht]
\begin{tabular}{c}
\includegraphics[height=6cm,clip=True]{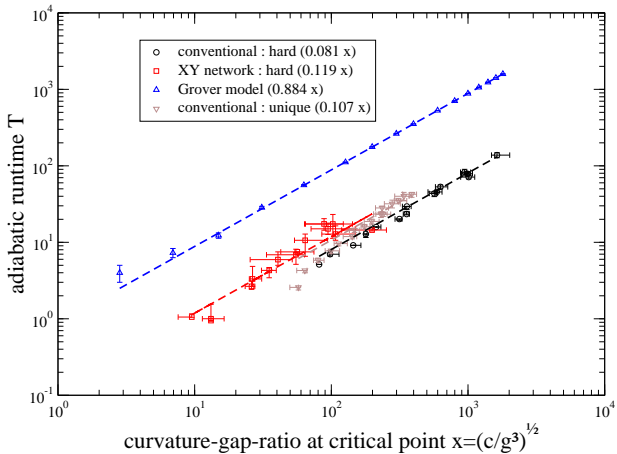}
\end{tabular}
\caption{\label{Fscaling}
[Color Online]
Scaling of the adiabatic runtime from figure \ref{Fmintime} versus 
$\sqrt{c_{\rm min}/g_{\rm min}^3}$ for different adiabatic algorithms.
All data sets agree nicely with the relation (dashed fit lines) predicted by
equation (\ref{Escaling}). 
With the exception of the Grover algorithm, each symbol corresponds to the median 
of 100 random instances with a unique satisfying agreement without constraints (brown) on 
the number of clauses or with $m \le 2/3 n$ clauses (black and red).
Values for the linear fits (dashed lines) are given in brackets.
Horizontal error bars give the 99\% confidence interval on the median, whereas vertical
error bars result from the finite precision when determining the necessary runtime by
repeated integration of the Schr\"odinger equation.
The data sets for Grover interpolation stem from \cite{schaller2006b},
where a different success probability of $P_{\rm final}^{\rm G}=0.75$
had been demanded. 
}
\end{figure}

We have numerically calculated the curvature and value of the minimum 
gap for interesting adiabatic optimization problems (see the following 
subsections) and have then related the adiabatic runtime $T$ with its
estimator $\sqrt{c_{\rm min}/g_{\rm min}^3}$. 
Our numerical experiments also confirm this scaling law quite nicely, see figure 
\ref{Fscaling}.

\section{Quadratic Spin Hamiltonians}\label{Squadratic}

Our conjecture (subsection \ref{SSconjecture}) 
can be exploited to improve the search for ground states of 
quite general quadratic Hamiltonians.
It is well known that the ground state of a frustrated
anti-ferromagnet under the influence of a local field as described by
the Hamiltonian 
\bea
\label{Equadratic}
H_{\rm F} = \sum_{i=1}^n f_i \sigma^z_i + \sum_{i,j=1}^n f_{ij}
\sigma^z_i \sigma^z_j
\eea
is in general extremely (NP-)hard
to find \footnote{From the classical point of view, finding the ground
  state of (\ref{Equadratic}) corresponds to finding the $n$-bit
  bit-string $z_1 \ldots z_n$ for which \mbox{$E = \sum_{i=1}^n f_i
    (1-2 z_i) + \sum_{i,j} f_{ij} (1 - 2 z_i) (1 - 2 z_j)$} has the
  minimum value. The quadratic form $f_{ij}$ can always be
  made positive definite without changing its value on the binary
  numbers by adding terms of the form $z_i^2 - z_i$. Although the
  minimum of a positive definite quadratic form in $n$ dimensions can
  easily be found in the set of real numbers with a standard
  minimization algorithm, is must be kept in mind that the above-posed
  problem implies the side constraints that the numbers have to be
  binaries. These side constraints have to be expressed by
  higher-order polynomials [such as e.g., $\sum_i x_i^2(1-x_i)^2$],
  which lead to the existence of many local minima and classical
  minimization algorithms will have difficulties to find the global
  minimum.}. 
In above equation, the coefficients $f_i$ and $f_{ij}=f_{ji}$ (with $f_{ii} = 0$)
denote the local field and the interaction topology, respectively.
More technically speaking, deciding whether the minimum energy of 
(\ref{Equadratic}) lies below a certain
threshold belongs to the class NP, i.e., 
it can be verified for a given solution with polynomial effort -- whereas actually
finding the solution can be much (e.g., exponentially) harder. 
For Exact Cover 3, this decision problem is even 
NP-complete \cite{farhi}, which implies that all other problems in NP
(i.e., decision versions of factoring, traveling salesmen problem, etc.)
can be mapped to it with using only polynomial overhead.
There is no classical algorithm known that delivers a solution using
only polynomial resources in $n$ and the existence of such an
algorithm would disprove the commonly accepted conjecture $P\neq NP$,
which is of vital importance in classical complexity theory. 

The discovery of a polynomial-time algorithm would have enormous consequences:
The (worst-case) performance of all known classical algorithms scales
exponentially in the problem size $n$ and already the solution of
relatively small problems becomes classically impossible for all
practical purposes.
This is for example exploited in encryption protocols, see also section
\ref{Snumberfactorization} and section \ref{Stea} in the appendix. 

\subsection{$XYZ$-Network}\label{Sxyz}

In order to find a good adiabatic quantum algorithm 
to reach the ground state of (\ref{Equadratic}), we want to use
the insight from the previous sections and try to construct an initial
Hamiltonian $H_{\rm I}$ such that the interpolation from $H_{\rm I}$ to 
$H_{\rm F}$ corresponds to a phase transition of second or higher
order. 
Since broken or restored symmetries are typically (though not always)
associated with second-order phase transitions, we shall demand that
$H_{\rm I}$ shares a symmetry with $H_{\rm F}$ and that this symmetry
is broken in the initial ground state but restored in the final
solution state. 
Furthermore, $H_{\rm I}$ should retain the interaction topology
(i.e., bit structure) of $H_{\rm F}$. 

To this end, one can exploit the evident symmetry of the problem
Hamiltonian (\ref{Equadratic}) with respect to rotations around the
axis generated by 
\bea
\label{Ehamming}
\Sigma^z=\sum_{i=1}^n \sigma^z_i
\,,
\eea
which we will further-on denote as Hamming-weight operator, since its 
eigenvalues are associated with the Hamming weight of the corresponding
subspace.
An obvious example for a Hamiltonian where this rotational symmetry is
spontaneously broken is the $SO(3)$-invariant ferromagnetic
Hamiltonian 
\bea
\label{E-xyz}
H_{\rm I}^{xyz} 
&=& 
-\sum_{i,j=1}^n |f_{ij}| \,\f{\sigma}_i\cdot\f{\sigma}_j
\nn
&=& 
-\sum_{i,j=1}^n |f_{ij}| 
\left[
\sigma^x_i \sigma^x_j + \sigma^y_i \sigma^y_j + \sigma^z_i \sigma^z_j 
\right]
\,,
\eea
where we have used the same interaction topology $f_{ij}$
as in (\ref{Equadratic}). 
Clearly, this Hamiltonian is invariant under rotations around
arbitrary axes -- including (\ref{Ehamming}) -- whereas its $(n+1)$-fold degenerate
ground state singles out one specific direction, for example 
$\ket{\uparrow\uparrow\dots\uparrow}$. 

This degeneracy immediately poses the question of which of these ground
states should be used as the initial state.
In analogy to the Ising model, where only the subspace with even
bit-flip parity was relevant, this question can be answered by the
conserved quantity (\ref{Ehamming}).
Since the final solution state will have a fixed Hamming weight
$\Delta$, we should start in the corresponding subspace 
${\cal H}_\Delta=\{\ket{\psi}\,:\,\Sigma^z\ket{\psi}=\Delta\ket{\psi}\}$. 
Obviously, $\ket{\uparrow\uparrow\dots\uparrow}$ and 
$\ket{\downarrow\downarrow\dots\downarrow}$ correspond to 
$\Delta=n$ and $\Delta=-n$, respectively, and are not suitable 
(except in trivial cases). 
However, the state $\ket{S}=\ket{\to\to\dots\to}$ contains a
superposition of all Hamming weights and can be used to project out
an initial state with 
any desired value of $\Delta_k=2k-n$ {\em via} 
\bea
\label{projector}
\ket{{\rm in}_k} = \sqrt{\frac{2^n}{\left(
\begin{array}{c}
n\\
k
\end{array}
\right)
}}
\frac{1}{2n+1} \sum_{k=0}^{2n}
\exp\left\{2\pi i\,\frac{\Delta_k - {\Sigma}_z}{2n+1}\,k\right\}
\ket{S}
\,.
\eea
Up to normalization, above formula is just the Fourier decomposition of the Kronecker symbol
${\delta}(\Delta - {\Sigma}_z)$ and involves single-qubit
rotations only.
The state (\ref{projector}) can be prepared efficiently by different approaches 
such as for example projective measurements or adiabatic evolution \cite{childs2002}.
Alternatively, one could use relaxation with an appropriate energy penalty in
the Hamiltonian such as $({\Sigma}^z-\Delta)^2$.

Of course, in order to prepare the initial state $\ket{{\rm in}}$
correctly, one has to know the Hamming weight $\Delta$ of the solution 
bit-string -- and in most adiabatic algorithms, this knowledge will not
be available in advance.
But since there are for an $n$-bit problem only $n+1$ subspaces with  
different Hamming weights, a naive testing of all these possibilities
corresponds to a polynomial overhead only.
Moreover, for many problems such as {\em exact cover 3} discussed in
the next section, one can guess the rough value of $\Delta$ and
thereby limit the number of trials. 

\subsection{$XY$-Network}\label{Sxy}

So far, we considered the fully $SO(3)$ rotationally symmetric
ferromagnetic Hamiltonian (\ref{E-xyz}).
However, numerically we have found (at least for the 
{\em exact cover 3} problem) that the planar ferromagnetic model 
\bea
\label{Enewinitial}
H_{\rm I}^{xy} = - \sum_{i,j=1}^n |f_{ij}| 
\left[\sigma^x_i \sigma^x_j + \sigma^y_i \sigma^y_j \right]
\,,
\eea
with the same interaction topology $f_{ij}$ yields a better
performance of the adiabatic algorithm (in average). 
In contrast to the spherically symmetric Hamiltonian (\ref{E-xyz}),
which is invariant under rotations around an arbitrary axis, 
this planar Hamiltonian is merely axially symmetric, i.e., invariant
under rotations around the  $\Sigma^z$-axis, i.e., 
$\left[H_{\rm I}, \Sigma^z\right]=0$.

Our conjecture is that for cases where the $H_{\rm I}^{xyz}$ initial Hamiltonian
performs significantly worse, not only $\Sigma^z$ is conserved exactly
but -- since this initial Hamiltonian commutes with 
\mbox{$\f{n}\cdot \f{\Sigma}=\f{n}\sum_{i} \f{\sigma}_i$} -- one might obtain
additional nearly conserved quantities 
via $\left[\f{n}\cdot\f{\Sigma}, H_{\rm F}\right]\approx 0$.
Since for an arbitrary vector $\f{n}$, the final ground state may
not have the symmetry, such nearly conserved quantities would
strongly hamper the evolution from the initial ground state to
the final one.
Of course, the probability to find such an axis $\f{n}$ is greatly 
reduced for the $H_{\rm I}^{xy}$ initial Hamiltonians, 
which is consistent with our observation of increased performance.

Furthermore, the state $\ket{S}=\ket{\to\to\dots\to}$ 
(as well as arbitrary rotations of it around the $\Sigma^z$-axis) is
no longer the {\em exact} ground state of (\ref{Enewinitial}) in general.
These states  are separable (no entanglement) whereas the true ground
state of (\ref{Enewinitial}) is already entangled and cannot be given
analytically for the general case. 
Fortunately, the mean-field approximation typically works reasonably
well for a large number $n\gg1$ of pseudo-randomly connected spins and
hence the state $\ket{S}$ should provide a good {\em approximation} to
the ground state. 
Indeed, numerically \cite{arpack1998} we found that the state in
(\ref{projector}) 
possesses a large overlap (over $90\%$) with the exact ground state
(in the relevant subspace ${\cal H}_\Delta$) and therefore provides a
good initial state for the computation. 
Thus the exact $SO(3)$ degeneracy of the ground state in the
$xyz$-network (\ref{E-xyz}) is replaced by an approximate $O(2)$
degeneracy (mean-field approximation) in the $xy$-network
(\ref{Enewinitial}). 
Nevertheless, both models imply a symmetry-restoring transition and
thus should be of second order. 
Note that interestingly, neither (\ref{E-xyz}) nor (\ref{Enewinitial}) 
are diagonal in the Hadamard basis, such that the arguments in \cite{mosca} 
do not directly apply.
In addition, these Hamiltonians differ strongly from projection operators (compare
the discussion in \cite{znidaric+horvat,farhi-fail}).

\section{Exact Cover 3}\label{Sexactcover}

A paradigmatic example for an NP-complete problem is 
{\em exact cover 3}: 
In this problem, one wants to find a bit-string 
$z_1 z_2\ldots z_{n-1} z_n$, where the 
$n$ bits $z_\alpha\in\{0,1\}$ must satisfy $m$ constraints (clauses). 
For general 3-satisfiability (3-SAT), each of these clauses involves
three bits $\alpha,\beta,\gamma\in\{1, \dots, n\}$ and for the
specific {\em exact cover 3}-problem every clause is defined by the
constraint  
\bea
\label{Eexactcover}
z_\alpha+z_\beta+z_\gamma=1\,,
\eea
which has to be satisfied for every triple $(\alpha,\beta,\gamma)$.
A Hamiltonian encoding the solution to the {\em exact cover 3}-problem
in its ground state can be defined by performing a sum over the
(positive semidefinite) single-clause penalties 
(compare also \cite{banuls2006a,quantum_phase}) 
\bea
H_{\rm F} = 
\sum_{c=1}^m \left(z_\alpha^c + z_\beta^c + z_\gamma^c -
\f{1}\right)^2
\,.
\eea
Note that this problem Hamiltonian is slightly different from the
original approach to {\em exact cover 3} \cite{farhi}, which assigns a
fixed energy penalty to each violated clause and thus involves
three-qubit interactions \cite{latorre2004a}, but has the same ground 
state (for satisfiable problems) -- only some of the excitation energies differ. 
The above ansatz has the advantage that two-qubit operations suffice
for its implementation, i.e., it is quadratic in the Pauli matrices.
Inserting $z_i=\frac{1}{2}\left[\f{1}-\sigma^z_i\right]$ one obtains
\bea
\label{Ehfnew}
H_{\rm F} = m \f{1} - \sum_i  \frac{n_i}{2} 
\sigma^z_i + \sum_{i,j} \frac{n_{ij}}{4} \sigma^z_i \sigma^z_j
\,,
\eea
where $n_i > 0$ denotes the number of
clauses involving the $i^{\rm th}$ bit and $n_{ij}\geq0$ the number of
clauses involving both bits $i$ and $j$.

For {\em exact cover 3} our initial Hamiltonian (\ref{Enewinitial})
reads  
\bea
\label{Ehinew}
H_{\rm I}^{xy} = - \sum_{i,j=1}^n \frac{n_{ij}}{4} 
\left[\sigma^x_i \sigma^x_j + \sigma^y_i \sigma^y_j \right]
\,.
\eea
Note that due to the relations 
\mbox{$2 n_i = \sum_j n_{ij}$} and 
\mbox{$3 m = \sum_i n_i$},  the two-bit matrix $n_{ij}$ of the final
Hamiltonian defines an {\em exact cover 3}-problem completely. 

In contrast, the original approach \cite{farhi} employed the straight 
interpolation scheme with the initial Hamiltonian
\bea
\label{Ehiold}
H_{\rm I}^{\rm conventional} = 
\sum_i \frac{n_i}{2} \left[\f{1}-\sigma^x_i\right]
\,,
\eea
which has the unique ground state (\ref{Etotalsup}) and no broken 
symmetry.  
In addition, it only contains the single-bit structure of the final
Hamiltonian.

The Schr\"odinger equation is invariant under simultaneous transformations of 
time and energy, such that we need to compare the energy scales of our modified approach
with the energy scales of the conventional ansatz \cite{das2003}.
The maximum energy spread of the quadratic final Hamiltonian (\ref{Ehfnew}) is 
$\Delta E_{\rm max}=4m$, which is only a factor of four larger than the conventional final Hamiltonian
\cite{farhi}.
Likewise, the maximum energy of our initial Hamiltonian (\ref{Ehinew}) can be upper
bounded by $\Delta E_{\rm max}\le 6m$ (the same bound would apply for the XYZ-network), 
which is only a factor of two larger than 
$E_{\rm max} =  3 m$ for the conventional initial Hamiltonian (\ref{Ehiold}).
Therefore, the energy resources required by our modified algorithm are at most a factor of four 
larger than in the conventional approach.

Regarding the unknown Hamming weight of the solution, 
we found for {\em exact cover 3} that it is typically situated 
around $\Delta\approx n/3$, which drastically reduces the
polynomial overhead generated by trying every possible value of
$\Delta$, see Sec.~\ref{Squadratic}. 
\subsection{Hard exact cover 3 problems}

In the following, we will numerically compare the effect of choosing
the initial Hamiltonians (\ref{Ehinew}) or (\ref{Ehiold}) on the performance of adiabatic
quantum computation. 
Of course, this performance will depend on the selection of clauses in
general. 
In order to select sufficiently hard problems, we only consider 
{\em exact cover 3} problems with just an unique satisfying agreement, as
was also done in the original study \cite{farhi}. 
However, there is evidence that this restriction is not sufficient yet
for ensuring the highest computational complexity.
For general 3-SAT, the number of instances with a unique satisfying agreement 
in combination with few clauses becomes exponentially rare among all instances 
with a unique satisfying agreement \cite{znidaric}.
Also for {\em exact cover 3} random problem instances with a unique satisfying 
agreement include many simple sub-problems, which have to be sorted out in order 
to find hard instances \cite{young2008a}.
Classically, the transition \cite{kalapala} from satisfiable to
non-satisfiable problems for {\em exact cover 3} suggests that
problems with rather few clauses ($m \approx 0.62 n$) have the highest
computational complexity. 

In order to select problems near that classical phase transition in the small
qubit range that is accessible to us, we have randomly generated
two sets of 100 problem instances for each qubit number $n$:
One problem set included only {\em exact cover 3} instances that have
a unique satisfying solution and an arbitrary number of clauses 
(as considered in \cite{farhi}). 
Another problem class did include {\em exact cover 3} instances with a
unique solution and few clauses, specifically $m \le {\rm round}(2n/3)$.
Note that this restriction explains the slight triple clustering of datapoints
around qubit numbers divisible by 3 in figures \ref{Fmintime}, \ref{Fvalue},
\ref{Fderivative}, \ref{Fcurvature}, and \ref{Fgapval}.
Problem instances were generated similarly to \cite{farhi} by successively adding
random clauses until there was only one solution left.
In the first problem set (analogous to \cite{farhi}) the result was discarded 
(and the procedure started over) whenever the problem became unsatisfiable.
In the second problem set we discarded the problem whenever either it
became unsatisfiable or when the number of clauses exceeded the boundary 
$m \le {\rm round}(2n/3)$.
Note that generating the latter problem set took a lot more effort than generating
the first, which indicates that problems with a unique solution but few constraints 
are quite rare among all problems with a unique satisfying agreement -- at least 
for the small qubit range accessible to us.

\subsection{Adiabatic Runtime}

\begin{figure}[ht]
\begin{tabular}{c}
\includegraphics[height=6cm,clip=True]{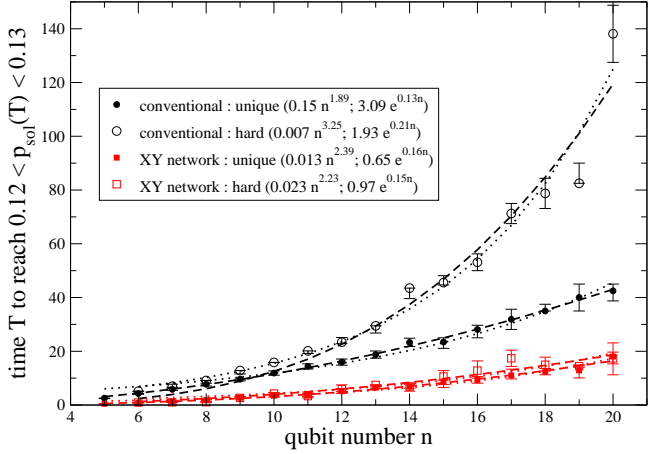}
\end{tabular}
\caption{\label{Fmintime}
[Color Online]
Runtime necessary to yield a final ground state fidelity between 0.12
and 0.13. 
In all cases the new algorithm (orange) performed in average better 
than the conventional approach (black).
Hollow symbols correspond to especially hard instances of the 
{\em exact cover 3} problem that do not only have a unique solution
but also few clauses, in this case $m \le {\rm round}(2n/3)$. 
(This leads to a triple clustering of the hollow symbols around qubit numbers 
divisible by 3.)
Dashed lines show polynomial fits, whereas dotted lines display exponential fits
(values given in brackets). 
}
\end{figure}

Since analytic solutions are unfortunately not available, we have to
compare the performance of the conventional approach (\ref{Ehiold})
with our modified proposal (\ref{Ehinew}) numerically. 
To this end, we first compute the adiabatic runtime $T$ that is 
necessary to obtain a fixed final fidelity of $1/8$ for both schemes,
see also \cite{quantum_phase}. 
In the conventional approach (\ref{Ehiold}), the quantum state was 
initialized with the state $\ket{S}$ in (\ref{Etotalsup}), 
whereas it has been initialized with the normalized projection
(\ref{projector}) of $\ket{S}$ onto the correct Hamming subspace 
in our scheme. 
For a range up to 20 qubits, the full Schr\"odinger equation had been
integrated using a fourth order Runge-Kutta scheme \cite{press1994}
with an adaptive step-size and varying run-times $T$, until an
acceptable success probability in the final state was found, see 
figure \ref{Fmintime}. 
A considerably improved algorithmic performance was found --
especially for the hard {\em exact cover 3} problems 
that have rather few clauses.
For these hard problems, the conventional scheme performed in average
significantly worse, whereas the new scheme did perform approximately
similar on both problem classes. 

However, it should be emphasized that the performance on a specific
problem may deviate significantly from the average behavior:
For example, we have also found some problems where the conventional
scheme performed better than the new algorithm.
The median plotted in figure \ref{Fmintime} for reasons of visibility
of error bars and compatibility with \cite{farhi} is hardly sensitive
to such rare instances. 
The worst case runtime we find (not shown) is not even a monotonously growing
function of the number of qubits.
For a given problem, it is therefore always practical to apply both quantum algorithms,
since the solution can be tested in polynomial time.

\subsection{Spectral Properties}

\begin{figure}[ht]
\begin{tabular}{c}
\includegraphics[height=6cm,clip=True]{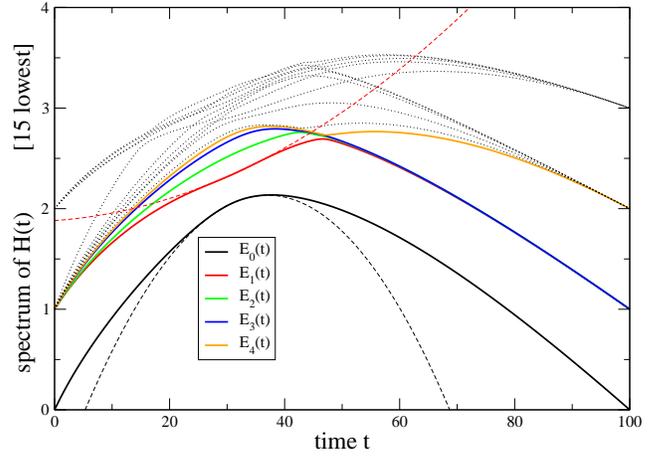}
\end{tabular}
\caption{\label{Fec3_example}
[Color Online]
Lower part of the spectrum for a typical {\em exact cover 3} instance
with 10 qubits. 
After the position of the minimum gap was determined by a minimization 
algorithm, parabolae (dashed lines) were fitted to the lowest two
eigenvalues. 
}
\end{figure}

\begin{figure}[ht]
\begin{tabular}{c}
\includegraphics[height=6cm,clip=True]{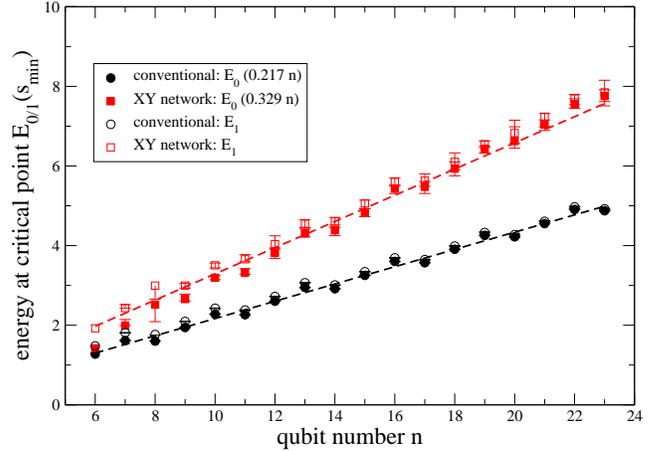}
\end{tabular}
\caption{\label{Fvalue}
[Color Online]
Value of the energies of ground state (solid symbols) and the first
excited state  
(hollow) at the position of the minimum gap for the conventional interpolation
scheme (black) and the new approach (red) {\em versus} the number of qubits.
Each symbol corresponds to the median of 100 random instances with a unique 
satisfying agreement and $m \le 2/3 n$ clauses and error bars give the 99\% 
confidence interval on the median.
Dashed lines display linear fits (fit parameters given in the legend in brackets)
to the ground state energies and demonstrate
that the modified algorithm has only a moderate increase in the ground state energy.
}
\end{figure}

\begin{figure}[ht]
\begin{tabular}{c}
\includegraphics[height=6cm,clip=True]{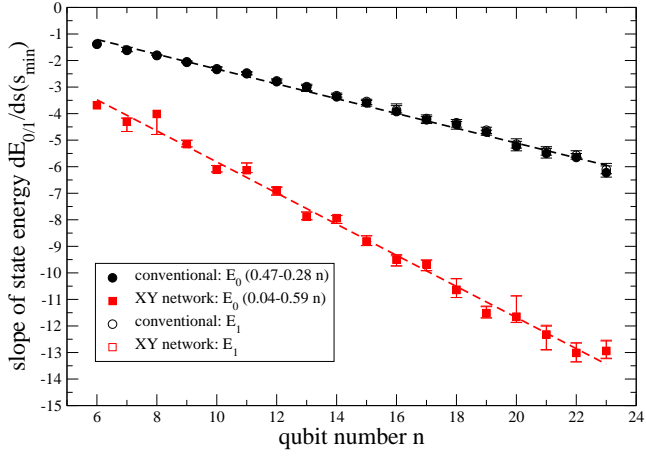}
\end{tabular}
\caption{\label{Fderivative}
[Color Online]
Value of the derivative of the lowest two energies at the position of 
the minimum gap. 
Invisibility of the hollow symbols demonstrates that an extremal point 
of the fundamental gap has been found with high accuracy.
Color coding and statistics are analogous to figure \ref{Fvalue}.
}
\end{figure}
 
\begin{figure}[ht]
\begin{tabular}{c}
\includegraphics[height=6cm,clip=True]{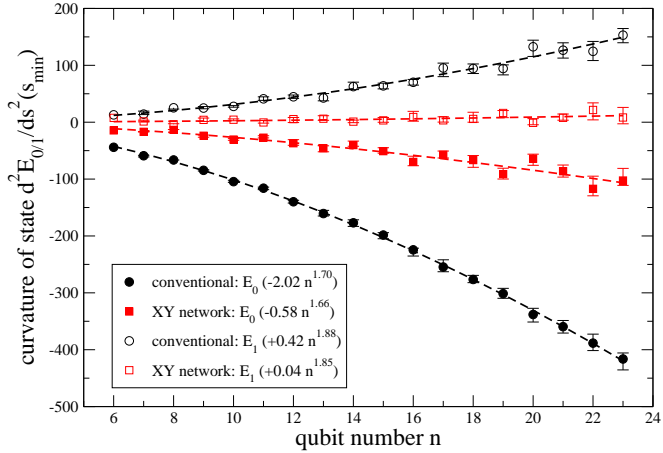}
\end{tabular}
\caption{\label{Fcurvature}
[Color Online]
Value of the curvature of the lowest two energies at the position of 
the minimum gap. 
Dashed lines represent polymial fits (parameters shown in brackets).
Color coding and statistics are analogous to figure \ref{Fvalue}.
}
\end{figure}

In order to clarify the origin of the speed-up found for our scheme, 
we have also analyzed the lower part of the spectrum with the ARPACK
package \cite{arpack1998}: 
After determining the position of the critical point with a
minimization algorithm 
\footnote{Of course, for the determination of the eigen-energies of the
new approach we did only consider the relevant Hamming subspace.}
applied to $g(s)=E_1(s)-E_0(s)$, a parabola was fitted to $E_0(s)$ and
to $E_1(s)$, see figure \ref{Fec3_example}. 
Since determining the lower part of the spectrum at a defined position
is more efficient than integrating the full Schr\"odinger equation, we
could extend our data range up to 23 qubits. 
In addition only the hard problem set with few clauses was considered
here. 
The parameters of this parabola have been averaged over the 100 
problem instances and are displayed in figures \ref{Fvalue},
\ref{Fderivative}, and \ref{Fcurvature}, respectively. 
As one would expect, the scaling of the critical ground state energy
is roughly linear for both algorithms, just the slope differs, see
figure \ref{Fvalue}. 
It is also visible from figure \ref{Fvalue} that the gap becomes
smaller with increasing system size (as expected) and that the new
approach has a larger minimum gap. 
The similar qualitative behavior holds also true for the average
slope of the ground state energy at the critical point, see figure
\ref{Fderivative}. 
However, for the curvature of the ground state energy -- which is a  
direct marker for the order of the quantum phase transition, compare
subsection \ref{SSising} --  we obtain significant differences between the
two algorithms. 
Whereas for the conventional algorithm the magnitude of the curvature
of the ground state (and the first excited state) at the critical
point increases strongly, for the new algorithm this scaling is
drastically reduced -- though still existent. 
This provides some evidence that by choosing the initial Hamiltonian of the form
(\ref{Ehinew}) with an approximately broken symmetry and retaining the
two-bit structure of the final Hamiltonian one may indeed improve
the order of the phase transition for {\em exact cover 3}. 

\subsection{Behavior of the fundamental gap}

\begin{figure}[ht]
\begin{tabular}{c}
\includegraphics[height=6cm,clip=True]{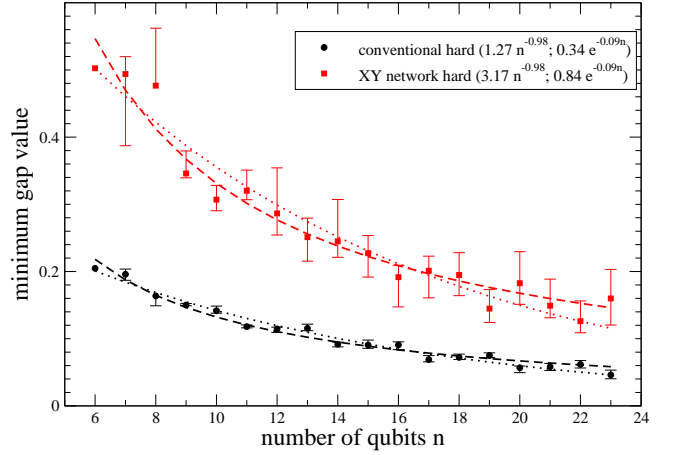}
\end{tabular}
\caption{\label{Fgapval}
[Color Online]
Scaling of the value of the minimum gap versus the number of qubits.
As one would expect, both scaling behaviors show a decrease with the
system size.
The vital question whether this scaling is polynomial (dashed fit lines) 
or exponential (dotted fit lines) cannot be answered with the limited numerical data.
The minimum gap of our modified scheme is a factor of about 2.5 larger than in
the conventional approach.
Color coding and statistics are analogous to figure \ref{Fvalue}.
}
\end{figure}

\begin{figure}[ht]
\begin{tabular}{c}
\includegraphics[height=6cm,clip=True]{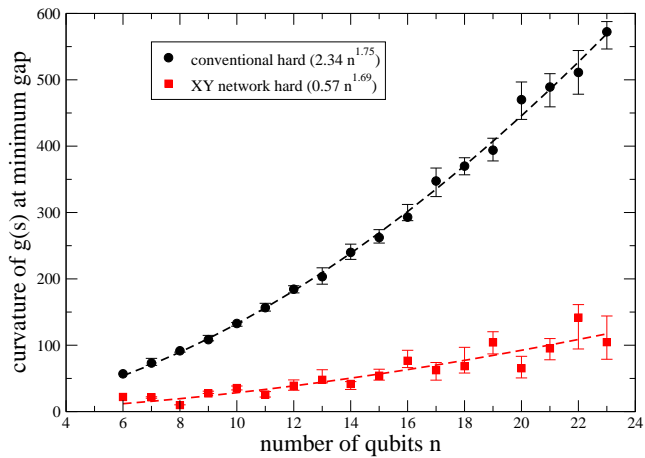}
\end{tabular}
\caption{\label{Fgapcrv}
[Color Online]
Scaling of the fundamental gap curvature at the minimum gap versus the number 
of qubits.
The curvature of our modified algorithm grows much slower as a function of $n$ than
with the conventional approach.
Color coding and statistics are analogous to figure \ref{Fvalue}.
}
\end{figure}

Although the absolute value of the minimum gap is larger 
in our new algorithm than in the conventional approach (see figure
\ref{Fgapval}), the ratio of the two values does not seem entirely sufficient
for explaining the strong differences in the adiabatic runtime in
figure \ref{Fmintime}. 
Recalling the results of section \ref{Sruntime}, we see that these
results are compatible with the different behavior of the gap
curvature, which can be deduced from figure \ref{Fcurvature} and is shown
explicitly in figure \ref{Fgapcrv}.
%
This is evidence that the adiabatic runtime can be positively influenced not 
only by the minimum fundamental gap alone but also by its curvature at the
critical point.
Note however that this is not too surprising, since for a smooth fundamental 
energy gap with a single minimum and that is bounded from below 
initially and finally, purely geometric arguments suggest that 
minimum gap and curvature at the minimum gap are related.

\subsection{Entropy of Entanglement}

Apart from the spectral properties of the Hamiltonian, entanglement is
another very useful concept for the understanding of quantum phase
transitions as well as quantum computing \cite{orus2004a,nielsen2000}.
If a quantum system is in a pure state $\rho = \ket{\Psi}\bra{\Psi}$,
the entanglement between two subsystems can be quantified by the
von-Neumann entropy 
\bea
\label{Eentropy}
S_{\rm E}
&=& 
S(\rho_1) = - {\rm Tr}_1\left\{\rho_1\log_2\rho_1\right\}
\nn 
&=& 
S(\rho_2) = - {\rm Tr}_2\left\{\rho_2\log_2\rho_2\right\}
\,,
\eea
of the reduced density matrix of either subsystem
\bea
\rho_1 = {\rm Tr}_2\left\{\rho\right\}
\,,\;
\rho_2 = {\rm Tr}_1\left\{\rho\right\}
\,,
\eea
where ${\rm Tr}_1$ and ${\rm Tr}_2$ denote the average over the
degrees of freedom of subsystem 1 and 2, respectively. 
The quantity $S_{\rm E}$ is called entanglement entropy and has the
property to vanish if and only if system and reservoir are not
entangled, i.e., 
$S_E=0\Leftrightarrow\ket{\Psi}=\ket{\Psi_1}\otimes\ket{\Psi_2}$. 
It is also bounded by the smaller number of qubits in either 
subsystem $S_E\leq{\rm min}\{n_1,n_2\}$. 

For general (sequential and adiabatic) quantum algorithms, it can be
shown that an exponential speed-up is only possible if the
entanglement $S_E$ contained in the system (at some point of the
calculation) grows sufficiently fast with system size \cite{vidal}. 
On the other hand, a nonanalytic (e.g., diverging) behaviour of the
entanglement $S_{\rm E}$ at the critical
point (in the infinite size limit) is also a typical feature of quantum
phase transitions.
Let us illustrate these results by means of a few examples:
The Ising model discussed in subsection \ref{SSising} (which is analytically 
solvable) displays a logarithmic scaling of the 
entanglement entropy near the critical point:
The ground state entanglement entropy of a block of $L$ spins 
scales mildly as $S_{\rm E}\sim\ln L$, see e.g.,
\cite{cincio2007a} and references therein. 
In contrast to spin systems in higher spatial dimensions with a
typical power-law scaling $S_E\sim n^p$, for example, this logarithmic
growth rate would not be sufficient for an exponential speed-up.
Shor's algorithm contains order-finding \cite{shor}, where the entanglement 
entropy between source and target register scales approximately
linearly with with the system size $n$ \cite{orus2004a}.
The same scaling has been found (numerically) for adiabatic algorithms 
for {\em exact cover 3} by calculating the entanglement entropy for the first
$n/2$ spins with the rest \cite{orus2004a}.
In comparison, the adiabatic Grover algorithm (with the same bipartition chosen)
does also assume a maximum value of the entanglement entropy at the critical point -- but
this value remains constrained by one for all system sizes
\cite{orus2004a}. 

Here, we have numerically analyzed the behavior of the
entanglement entropy of the instantaneous ground state.
Imagining the $n$ qubits as being lined up in a chain, we calculated 
the entanglement entropy between the subsystems formed by the first
$n_1$ qubits and the remaining $n_2=n-n_1$ qubits.  
Numerically, the entanglement entropy was found by calculating the 
instantaneous ground state $\ket{\Psi_0(t)}$ of the system
\cite{arpack1998} and then determining the reduced density matrix
$\rho_1$. 
Due to $S(\rho_1) = S(\rho_2)$ the largest reduced density matrices
for $n=10$ qubits have $N^2=32^2$ entries and can therefore be
directly diagonalized \cite{press1994}. 
The $n-1$ values for the entanglement entropy resulting from the
different partitions of an $n$-bit chain have simply been averaged.
This resulting average was then again averaged over 100 (hard) problem
instances of {\em exact cover 3}. 

\begin{figure}[ht]
\begin{tabular}{c}
\includegraphics[height=6cm,clip=True]{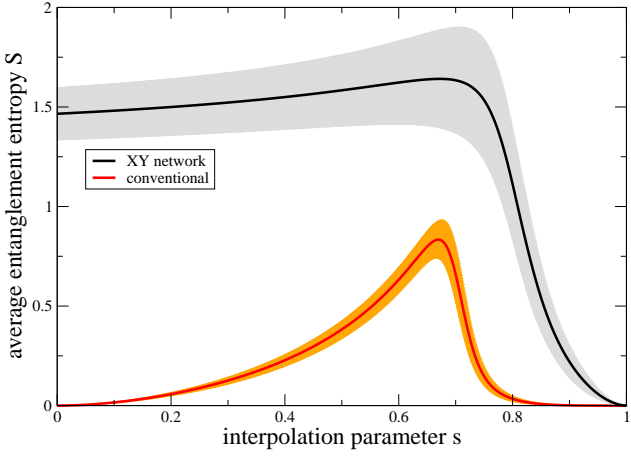}
\end{tabular}
\caption{\label{Fentropy}
[Color Online]
Behavior of the average entanglement entropy for a system of 10
qubits. 
The shaded regions correspond to one standard deviation and the lines
show the average over 100 problem instances.
}
\end{figure}

As one might have expected, the curves for the conventional scheme and
our proposal differ strongly, see figure \ref{Fentropy}.
For the conventional scheme, the entanglement entropy vanishes at the
beginning and at the end of the calculation and possesses a pronounced
peak at the critical point.
In our algorithm, this peak is smeared out and we already start off
with a relatively large entropy -- the initial state in
(\ref{projector}) is entangled, for a discussion of the required resources see 
subsection \ref{Sxyz}.
Moreover, the entropy is significantly larger in our algorithm
throughout the interpolation. 
Both observations (i.e., the fact the the entropy is larger and varies
slower) could be interpreted as indications for the increased
algorithmic performance -- but one should bear in mind that the exact
relation between the entanglement entropy and the achievable speed-up
is not fully understood yet. 

Increasing the number of qubits, we do also see the approximately  
linear scaling of the maximum entanglement entropy
(in the qubit range that is accessible to us) 
observed in the literature \cite{banuls2006a}. 

\section{Number Factorization}\label{Snumberfactorization}

Apart from {\em exact cover 3}, there are many other problems whose
solution can be encoded in the ground state of a quadratic Hamiltonian
in (\ref{Equadratic}). 
As one example, let us discuss the factoring problem (which is in NP,
but not believed to be NP-complete), a further example is outlined in the appendix
\ref{Stea}. 

Given a large number $\omega$, it is in general very difficult to 
answer the question which numbers $a$ and $b$ fulfill $\omega=ab$.  
The product of a $k$-bit number $a$ and another ($n-k$)-bit number $b$
has either ($n-1$) or $n$ bits.
In the following, we assume the latter case (the other option can
be easily adapted by using $\omega_1 = 0$) 
and restrict our considerations to a
Hilbert space with $n$ qubits where the basis states can be written as  
$\ket{\Psi}=\ket{a_1}\dots\ket{a_k}\ket{b_1}\dots\ket{b_{n-k}}$.
Of course, even for problems with a unique solution (e.g., bi-primes
with $k\neq n/2$), the partition $(k, n-k)$ will usually not be known
in advance. 
However, the associated worst-case overhead of trying every
possible value of $k$ from $n/2$ to $n$ grows only linearly as a 
function of $n$. 
A naive Hamiltonian encoding the solution as the ground state is then
readily given by 
\bea
H_{\rm F}^{(1)} = (\omega - \hat{a} \hat{b})^2\,,
\eea
where 
\bea
\hat{a} &=& \sum_{\ell=1}^k \hat{a}_\ell 2^{k-\ell} 
\quad:\quad 
\hat{a}_\ell = \frac{1}{2}(\f{1}-\sigma^z_\ell)\,,\nn
\hat{b} &=& \sum_{\ell=1}^{n-k} \hat{b}_\ell 2^{n-k-\ell}
\quad:\quad 
\hat{b}_\ell = \frac{1}{2}(\f{1}-\sigma^z_{k+\ell})
\,.
\eea
Although it operates on a Hilbert space with dimension $n$ and uses 
interactions between at most four qubits, the above Hamiltonian has
the disadvantage that its couplings (operator pre-factors) cover an
exponential range and thus its spectral range 
(ratio of largest and smallest eigenvalue) increases exponentially 
with the number of qubits $n$.   
In an experiment, an exponential fine-tuning of the couplings will
probably be infeasible for the interesting case of large $n$.
In addition, an exponentially increasing spectral width will be 
hard to realize without using exponential resources.

This scaling of the spectral width could be improved by using the
Hamming distance as a penalty in the Hamiltonian
\bea
H_{\rm F}^{(2)} = \sum_{\ell=1}^n \left[\omega_\ell 
+ (-1)^{\omega_\ell} \left(\hat{a}\hat{b}\right)_\ell \right]
\,,
\eea
where $\omega_\ell$ denotes the $\ell$-th bit of $\omega$, since here,
the spectral width grows only linearly with the number of qubits. 
However, the computation of the $\ell$-th bit of the product
$\hat{a}\hat{b}$ requires simultaneous interactions of multiple
qubits. 

In the following, we construct a Hamiltonian that has at most 
two-qubit interactions and simultaneously a spectral
width that scales only polynomially with the number of qubits.
In addition, we demand that the local coupling constants should cover a
finite range only.

\subsection{Factorization Equations}

To introduce our approach, let us consider as a generic example the
case \mbox{$n=10$} and \mbox{$k=6$}.  
Then, the usual method of multiplying two binary numbers can be
decomposed into $n$ equations according to Table \ref{Tfirst} such as 
\bea
\label{Efactorization_equation}
a_1 b_4 + a_2 b_3 + a_3 b_2 + a_4 b_1 + z_{65} + z_{75} &&\nn
 - \left( \omega_5 + 2 z_{54} + 4 z_{53}\right) &=& 0
\,,
\eea
where the binary variables $z_{i>j}$ represent the carries from
columns $i$ towards $j$ 
(the above equation corresponds to the fifth column). 
The complete set of equations like (\ref{Efactorization_equation}) 
are also called factorization equations, where
a linear (plus logarithmic corrections) amount of
carry variables $z_{ij}$ is needed, see \cite{burges} for a detailed
discussion.  
In an adiabatic quantum computer, these carry variables would have to
be represented by additional qubits.

\begin{table}[ht]
\begin{tabular}{c|c|c|c|c|c|c|c|c|c}
&	&	&	& $a_1$	& $a_2$	& $a_3$	& $a_4$	& $a_5$ & $a_6$\\
&	&	&	&     	&     	& $b_1$	& $b_2$ & $b_3$ & $b_4$\\
\hline
&	&	&
&$a_1b_4$&$a_2b_4$&$a_3b_4$&$a_4b_4$&$a_5b_4$&$a_6b_4$\\ 
&	&	&$a_1b_3$&$a_2b_3$&$a_3b_3$&$a_4b_3$&$a_5b_3$&$a_6b_3$&\\
&	& $a_1b_2$ & $a_2b_2$ & $a_3b_2$ & $a_4b_2$ & $a_5b_2$&$a_6b_2$&\\
& $a_1b_1$ & $a_2b_1$ & $a_3b_1$ & $a_4b_1$ & $a_5b_1$&$a_6b_1$&&\\
\hline
$z_{21}$&$z_{32}$&$z_{43}$&$z_{54}$&$z_{65}$&$z_{76}$&$z_{87}$&$z_{98}$&&\\
$z_{31}$&$z_{42}$&$z_{53}$&$z_{64}$&$z_{75}$&$z_{86}$&&&&\\
\hline
$\omega_1$ & $\omega_2$ & $\omega_3$ & $\omega_4$ & $\omega_5$ &
$\omega_6$ & $\omega_7$ & $\omega_8$ & $\omega_9$ & $\omega_{10}$
\end{tabular}
\caption{\label{Tfirst}
Conventional method for multiplying the \mbox{$k=6$} bit number $a$
and the \mbox{$n-k=4$} bit number $b$. The result $\omega=a b$ is
obtained by summing the partial products $a_i b_j$ and carry bits
$z_{ij}$ (from column $i$ to column $j$), which gives rise to $n$
factorization equations. Note that for this example, at most two carry
bits are necessary in each column.} 
\end{table}

Obviously, factoring the number $\omega$ corresponds to finding a
solution for the binary variables
$(a_1,\dots,a_k,b_1,\dots,b_{n-k},z_{ij})$ to the factoring equations.
A simple idea is to assign a positive penalty to each
violated factoring equation {\em via} 
\bea
H_{\rm F} = \sum\limits_{i=1}^n E_i^2
\,,
\ea
where $E_i=0$ expresses the $i$th factoring equation in normal form
(with zero on the right hand side).
Note however, that by naively using the square of the factorization
equations we have generated four-qubit-interactions in the Hamiltonian.
The maximum penalty originating from a single equation scales as
$\ord\{(n-k)^2\}$, such that the complete spectral width of
this Hamiltonian scales in the worst case as $\ord\{n^3\}$ (and so
does the number of required four-qubit interactions).
Although only a linear number of auxiliary qubits is needed, it
may be experimentally difficult to realize so many different
four-qubit interactions. 
In addition, our approach to generate higher-order phase transitions 
cannot be directly applied, since the Hamiltonian does not have the
standard form of (\ref{Equadratic}).

In order to obtain only two-qubit interactions, further modifications
are necessary:
Consider the case where only a single product of binary variables $A$
and $B$ is involved in each factoring equation
\bea
\label{Efactorization_sample}
E = A B + S = 0
\,,
\eea
where $A$ and $B$ are single bit variables and 
\mbox{$S=\sum_j \alpha_j c_j$}
is a sum of single bit variables with $\alpha_j \in \mathbb{Z}$.
Then, the naive penalty for violating the equation
(\ref{Efactorization_sample}), which would be given by 
$P_i = \left[A B + S\right]^2$, can be replaced by 
\bea
\label{Epenalty_new}
\tilde{P}_E = 
2\left[\frac{1}{2}\left(A + B - \frac{1}{2}\right)+S\right]^2- \frac{1}{8}
\eea
in the sense that both penalties vanish if and only if equation  
(\ref{Efactorization_sample}) is obeyed [which implies
\mbox{$(AB=0 \wedge S = 0) \vee (AB=1 \wedge S=-1)$}]
and are larger or equal to one otherwise.
This modified penalty has the advantage that it involves only
quadratic interactions between the qubits. 

The task one is left with is to rewrite the multiplication table 
with the help of further ancilla variables such that only
factorization equations of the type (\ref{Efactorization_sample}) 
occur. 
For the generic example, this is demonstrated in table \ref{Tsecond}.

\begin{table}[ht]
\begin{tabular}{c|c|c|c|c|c|c|c|c|c}
&	&	&	& $a_1$	& $a_2$	& $a_3$	& $a_4$	& $a_5$ & $a_6$\\
&	&	&	&     	&     	& $b_1$	& $b_2$ & $b_3$ & $b_4$\\
\hline
&	&	&	&0	&0	&0	&0	&0	&0\\
&	&	&
&$a_1b_4$&$a_2b_4$&$a_3b_4$&$a_4b_4$&$a_5b_4$&$a_6b_4$\\ 
&	&	&	&0	&0	&0	&0	&0	&0\\
\cline{4-10}
&	&	&0	&$S_{23}$&$S_{33}$&$S_{43}$&$S_{53}$&$S_{63}$&\\
&	&	&$a_1b_3$&$a_2b_3$&$a_3b_3$&$a_4b_3$&$a_5b_3$&$a_6b_3$&\\
&	&	&$z_{13}$&$z_{23}$&$z_{33}$&$z_{43}$&$z_{53}$&0&\\
\cline{3-9}
&	&$S_{12}$&$S_{22}$&$S_{32}$&$S_{42}$&$S_{52}$&$S_{62}$&&\\
&	&$a_1b_2$&$a_2b_2$&$a_3b_2$&$a_4b_2$&$a_5b_2$&$a_6b_2$&&\\
&	&$z_{12}$&$z_{22}$&$z_{32}$&$z_{42}$&$z_{52}$&0&&\\
\cline{2-8}
&$S_{11}$&$S_{21}$&$S_{31}$&$S_{41}$&$S_{51}$&$S_{61}$&&&\\
&$a_1b_1$&$a_2b_1$&$a_3b_1$&$a_4b_1$&$a_5b_1$&$a_6b_1$&&&\\
&$z_{11}$&$z_{21}$&$z_{31}$&$z_{41}$&$z_{51}$&0&&&\\
\hline
$\omega_1$ & $\omega_2$ & $\omega_3$ & $\omega_4$ & $\omega_5$ &
$\omega_6$ & $\omega_7$ & $\omega_8$ & $\omega_9$ & $\omega_{10}$
\end{tabular}
\caption{\label{Tsecond}
Multiplication table for multiplying the 6 bit number $a$ and the 4
bit number $b$ using only two-qubit interactions.
In each row, the first line contains the partial
product sums, the second line the partial products, and the last line
the carry variables. 
Note that the meaning of the indices for the auxiliary variables has
changed in comparison to table \ref{Tfirst}.
Apart from the boundaries, the binary variables have to fulfill 
\mbox{$a_i b_j + S_{ij} + z_{ij} = S_{i+1,j-1} + 2 z_{i-1,j}$}
(see the text for explanations). 
For each of the ($n-k-1=3$) horizontal lines that
have been inserted, $k=6$ partial product variables $S_{ij}$ and
($k-1=5$) carry variables $z_{ij}$ have to be included. 
The first partial product variable (second row) always vanishes.
Note also that the first partial product variable in each row is
equal to the carry variable in the row above, which has therefore been
removed.}
\end{table}

The general structure of the equations is for $1\le i \le k$ and 
$1\le j \le n-k$ is given by
\bea
\label{Egeneral}
a_i b_j + S_{ij} + z_{ij} = S_{i+1,j-1} + 2 z_{i-1,j}
\,,
\eea
where at the boundaries (marked by invalid indices) the equations are
closed by 
\bea
z_{0,j} &=& S_{1,j-1}\,,\nn
S_{i,0} &=& \omega_{i}\,,\nn
S_{k+1,j-1} &=& \omega_{k+j}\,,\nn
S_{i,n-k} &=& z_{i,n-k} = 0\,,\nn
z_{k, j} &=& 0\,,\nn
S_{1,n-k-1} &=& 0
\,.
\eea
With the suggested replacement in (\ref{Epenalty_new}), this leads to
the total penalty Hamiltonian 
\bea
\label{Ehamtwo}
H_{\rm F}^{(3)} &=& 
\sum_{i=1}^{k} \sum_{j=1}^{n-k}
\left\{2\left[ \frac{1}{2} \left(\hat{a}_i + \hat{b}_j - \frac{1}{2}\right)
+ \hat{S}_{ij} + \hat{z}_{ij}
\right.\right.
\nn
&& 
\left.\left. 
\phantom{\frac12}
-
\hat{S}_{i+1, j-1} - 2 \hat{z}_{i-1,j}\right]^2 - \frac{1}{8}\right\}
\,,
\eea
where all operators act on distinct single qubits (for example on a
two-dimensional lattice arrangement as 
\mbox{$\hat{S}_{ij}=(\f{1}-\sigma^z_{ij})/2$}).
The above Hamiltonian has the ground state (an appropriate qubit
ordering assumed)
\bea
\ket{\Psi_{\rm g}} = 
\ket{a_1 \dots a_k}\ket{b_1 \dots b_{n-k}}
\ket{\{S_{ij}\}}\ket{\{z_{ij}\}}
\,,
\eea
where $\omega=ab$ is the sought-after factorization. 
The number of necessary auxiliary variables can be calculated as
follows: 
In order to multiply a $k$-bit number $a$ and a $(n-k)$-bit number
$b$, at most $k(n-k)$ partial products are required.
For separating these into sums with a single product of two
binary variables, ($n-k-1$) horizontal lines have to be inserted, compare
table \ref{Tsecond}.
Each of these lines requires $k$ partial sum variables $S_{ij}$
(except in the second row, where the first $S_{1,n-k-1}$ vanishes) 
and ($k-1$) carry variables $z_{ij}$.
Note that the bottom ($k-1$) carry variables can be associated with
the vanishing top ($k-1$) carry variables.
Therefore, the total number of auxiliary variables is given by
$(2k-1)(n-k-1)-1$.
Together with the $n$ bits required for the factors $a$ and $b$, 
at most $n-1+(2k-1)(n-k-1)$ qubits are required for finding the $k$
and $n-k$ bit factors of an $n$-bit number.
On an adiabatic quantum computer, this implies a quadratic overhead
in the number of qubits. 
Since the number of variables in each factoring equation is always
smaller or equal to 6, the total number of quadratic interactions
scales with the number of equations, i.~e., quadratically in $n$.
For each equation in (\ref{Ehamtwo}), the largest possible penalty is
21, which leads to a quadratical scaling of the spectral width.
Note also that the necessary coupling strength between different
qubits ranges (independent of the problem size) from 1 to 8.

\subsection{Numerical Study}

Similar to satisfiability problems -- where those with a unique
solution are believed to belong to the hardest problems \cite{farhi} 
(both classically and in adiabatic quantum algorithms), bi-primes
(products of two prime numbers) possess a unique factorization and the
classical hardness of bi-prime factorization is used in many
cryptography protocols. 
Except in the case $k=n/2$ (where the solution becomes two-fold
degenerate due to $\omega=ab=ba$), the Hamiltonians have a unique
ground state. 
In our case, biprime factorization constitutes a well-defined (and
also classically relevant) problem class, for which the algorithmic
performance of an adiabatic quantum algorithm using (\ref{Ehamtwo}) as 
the problem Hamiltonian would be interesting.

In order to reduce the problem complexity of (\ref{Ehamtwo}), one can
use the fact that both prime factors are odd in nontrivial cases and
that there exists a minimum size of the prime factors to generate an
$n$-bit number, i.e., that both first and last bits of the prime
factors are set to 1, see table \ref{Tthird} for the generic example. 
Under these conditions, the first row generates $k-1$ nontrivial
equations that involve no product between different qubits at all. 
In these, the $S_{ij}$ variables can be eliminated without changing
the structure of the equations.
Together with the $(n-4)$ bits required to store the unknown bits of
the prime factors, one arrives at \mbox{$[2k(n-k-1) - 3]$} variables
to find the odd $(k,n-k)$ factors of the $n$-bit number $\omega$.
Finally, the two-qubit Hamiltonian can be cast into the form 
\bea
\label{Efactquad}
H_{\rm F} = h + \sum_{i=1}^{n_{\rm tot}} h_i \sigma_i^z 
+  2 \sum_{i=1}^{n_{\rm tot}}\sum_{j=i+1}^{n_{\rm tot}} h_{ij}
   \sigma_i^z \sigma_j^z
\,,
\eea
where $n_{\rm tot}$ denotes the total number of qubits and  
\mbox{$h,h_i,h_{ij}\in\mathbb R$} with $h_{ij}=h_{ji}$ and $h_{ii}=0$.
It is easy to see that the coefficients $h, h_i, h_{ij}$ can be
conveniently extracted from eqn. (\ref{Ehamtwo}) by computing traces of 
$H_{\rm F}^{(3)}$, $\sigma^z_i H_{\rm F}^{(3)}$, and 
$\sigma^z_i \sigma^z_j H_{\rm F}^{(3)}$, respectively.
We have analyzed the algorithmic performance for a linear
interpolation for three different initial Hamiltonians 
\bea
\label{Einithams}
H_{\rm I}^{\rm x} 
&=& 
\sum_{i=1}^{n_{\rm tot}}
\frac{1}{2}\left[\f{1}-\sigma_i^x\right]\,,
\nn
H_{\rm I}^{\rm xy} 
&=& 
\sum_{i,j=1}^{n_{\rm tot}} \frac{\abs{h_{ij}}}{4}
\left[\f{2} - \sigma_i^x\sigma_j^x - \sigma_i^y\sigma_j^y\right]\,,
\nn
H_{\rm I}^{\rm xyz} 
&=& 
\sum_{i,j=1}^{n_{\rm tot}} \frac{\abs{h_{ij}}}{4}
\left[\f{1} - 
\sigma_i^x\sigma_j^x - \sigma_i^y\sigma_j^y -
\sigma_i^z\sigma_j^z\right]
\,.
\eea
Note that the first Hamiltonian has the ground state $\ket{S}$ in
(\ref{Etotalsup}) and no apparent symmetry of the final Hamiltonian
is respected.
The latter two Hamiltonians commute with the Hamming-weight operator
(\ref{Ehamming}) and the same discussion as in section
\ref{Sexactcover} applies.
However, there are also some crucial differences:
\begin{itemize}
\item
In contrast to {\em exact cover 3}, the numbers $h_{ij}$ in the
quadratic decomposition (\ref{Efactquad}) are not necessarily
positive. 
\item
In addition, unlike {\em exact cover 3}, the two-bit interactions 
$h_{ij}$ do not completely determine the problem, such that the
Hamiltonians in (\ref{Einithams}) can probably be further improved.
\end{itemize}

\begin{table}[ht]
\begin{tabular}{c|c|c|c|c|c|c|c|c|c}
&	&	&	& 1	& $a_2$	& $a_3$	& $a_4$	& $a_5$ & 1\\
&	&	&	&     	&     	& 1	& $b_2$ & $b_3$ & 1\\
\cline{4-10}
&	&	&0	&1	&$a_{2}$&$a_{3}$&$a_{4}$&$a_{5}$&1\\
&	&	&$b_3$&$a_2b_3$&$a_3b_3$&$a_4b_3$&$a_5b_3$&$b_3$&\\
&	&	&$z_{13}$&$z_{23}$&$z_{33}$&$z_{43}$&$z_{53}$&0&\\
\cline{3-9}
&	&$S_{12}$&$S_{22}$&$S_{32}$&$S_{42}$&$S_{52}$&$S_{62}$&&\\
&	&$b_2$&$a_2b_2$&$a_3b_2$&$a_4b_2$&$a_5b_2$&$b_2$&&\\
&	&$z_{12}$&$z_{22}$&$z_{32}$&$z_{42}$&$z_{52}$&0&&\\
\cline{2-8}
&$S_{11}$&$S_{21}$&$S_{31}$&$S_{41}$&$S_{51}$&$S_{61}$&&&\\
&1&$a_2$&$a_3$&$a_4$&$a_5$&1&&&\\
&$z_{11}$&$z_{21}$&$z_{31}$&$z_{41}$&$z_{51}$&0&&&\\
\hline
$\omega_1$ & $\omega_2$ & $\omega_3$ & $\omega_4$ & $\omega_5$ &
$\omega_6$ & $\omega_7$ & $\omega_8$ & $\omega_9$ & 1
\end{tabular}
\caption{\label{Tthird}
Multiplication table for multiplying the odd 6 bit number $a$ and the
odd 4 bit number $b$. The ($k-1$) $S_{ij}$-variables in the top row of table
\ref{Tsecond} have been eliminated without changing the structure of
the equations.
One equation (top right) becomes trivial,
whereas (from top to bottom) $k-1+2(n-k-2)+k=2n-5$ equations emerge
that involve no products between different qubits at all.
}
\end{table}

\begin{figure}[ht]
\begin{tabular}{c}
\includegraphics[height=6cm,clip=True]{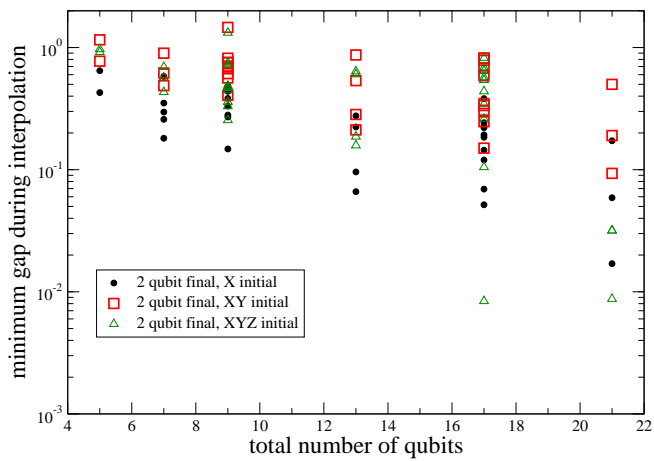}
\end{tabular}
\caption{\label{Fmingap}
[Color Online]
Minimum gap during adiabatic interpolation for number factorization
versus the total number of qubits used. 
For problems with a doubly
degenerate ground state, the minimum gap between the lowest and the second
excited state has been calculated.
Each data set has the same number of data points per column (some
symbols lie on top of each other). 
}
\end{figure}

Since we can with moderate effort simulate the evolution of systems
with $\ord(20)$ qubits, we can numerically access the $(n,n-k)$
factoring partitions displayed in table \ref{Tpartitions}. 
For the bi-primes within this range, we have determined the minimum 
fundamental gap during the linear interpolation \cite{arpack1998}, see
figure \ref{Fmingap}.
It is visible that again the $H_{\rm I}^{\rm xy}$ Hamiltonian is
superior to the other choices in (\ref{Einithams}) -- at least for the
small sample problems under consideration. 

Interestingly, the third Hamiltonian in (\ref{Einithams}) leads in
some cases to an even smaller minimum gap than with the conventional
choice.
We have also observed this for rare instances of EC3.
As discussed in subsection \ref{Sxy}, we conjecture additional nearly conserved quantities 
to hamper the adiabatic evolution towards the solution in case of the $H_{\rm I}^{\rm xyz}$ 
initial Hamiltonian.

\begin{table}[ht]
\begin{tabular}{c|c|c}
$n_{\rm tot}$ & bi-primes & partitions $(n,n-k)$\\
\hline
5  & 33;39 & (4,2);(4,2)\\
\hline
7  & 51;57;69;87;93 & (5,2);(5,2);(5,2);(5,2);(5,2)\\
\hline
9  & 25;35;49;111;123& (3,3);(3,3);(3,3);(6,2);(6,2)\\
   & 129;141;159;177;183 & (6,2);(6,2);(6,2);(6,2);(6,2)\\
\hline
13 & 55;65;77;91 & (4,3);(4,3);(4,3);(4,3)\\
\hline
17 & 85;95;115;119;133 & (5,3);(5,3);(5,3);(5,3);(5,3)\\
   & 145;155;161;203;217 & (5,3);(5,3);(5,3);(5,3);(5,3)\\
\hline
21 & 121;143;169 & (4,4);(4,4);(4,4)
\end{tabular}
\caption{\label{Tpartitions}
Bi-primes and partitions accessible with different numbers of qubits
(leftmost column). 
}
\end{table}

\section{Conclusions}

In summary, the analogy between adiabatic quantum algorithms and
quantum phase transitions yields new insight and facilitates a better 
understanding of both. 
For first-order transitions, the system has to tunnel through an 
energy barrier in order to stay in the ground state 
(cf.~Fig.~\ref{Fsketch_grover}), which indicates an exponential
scaling of the tunneling time with system size.   
Since such a tunneling barrier is absent for transition of second or
higher order (cf.~Fig.~\ref{Fsketch_ising}), we conjecture that
they are advantageous with respect to adiabatic quantum computation in 
the sense that they allow much shorter run-times.

Based on this physical intuition, we designed a modified adiabatic
quantum algorithm for the NP-complete problem {\em exact cover 3} 
in analogy to a symmetry-restoring quantum phase transition and found
numerically that it indeed yields an improved performance
in comparison with the conventional scheme. 
Even though the infinite-size limit of these adiabatic quantum algorithm
may not be as well-defined as that of the Ising model (probably they
should be classified somewhere between first and second order), the
spectral characteristics indicate that our alternative adiabatic quantum
algorithm has a higher order than the conventional scheme. 

As a closely related point, we observed that the adiabatic run-time 
(of constant-speed interpolations) is not just determined by the value
of the minimum gap but also by the curvature of the energy levels at
the critical point.  
In fact, the better average scaling behavior of our alternative algorithm can
probably be attributed mostly to the fact that the energy levels are
far less curved than in the conventional scheme.  

Apart from the spectral characteristics, entanglement is another
useful concept for quantum algorithms as well as quantum phase
transitions. 
We found that the ground-state entanglement entropy in our modified 
adiabatic quantum algorithm is much larger than that in the
conventional scheme, which might be connected with the advantages of
higher-order phase transitions. 
However, these advantages may also go along with some drawbacks:  
for the prototypical Grover and Ising models, we found that
second-order phase transitions seem to be more vulnerable to
decoherence than those of first order \cite{tiersch,mostame2007a}.  

Finally, even though the run-time scaling (polynomial versus
exponential) is not clear, the results obtained so far
strongly suggest that adiabatic quantum algorithm algorithm can solve
NP problems much faster than the Grover search routine (with a
quadratic speed-up). 
The methods proposed in this article can easily be applied to other
computationally relevant problems which can be encoded into the ground
states of Hamiltonians that are quadratic in the Pauli matrices -- 
such as factoring (see also the Appendix \ref{Stea}). 

\section*{Acknowledgements}

This work was partly supported by the Emmy Noether Programme of the
German Research Foundation (DFG) under grant No.~SCHU~1557/1-2. 
R.~S.~acknowledges fruitful discussions at the
Les Houches Summer School on Quantum Magnetism and the 
Banff (BIRS) workshop on ``Spin, Charge, and Topology'' 
(supported by PITP), 
and valuable conversations with G.~Volovik
(visits supported by EU-ULTI and ESF-COSLAB).
The authors are indebted to F.~Krauss for providing computational
resources and to E.~Farhi, J.~Goldstone, and R.~Plaga for fruitful
discussions. 

$^*$\,{\small\sf schaller@itp.physik.tu-berlin.de}

$^\dagger$\,{\small\sf schuetz@theo.physik.uni-due.de}

\section{Appendix: Encryption with TEA}\label{Stea}

Although factoring numbers is also used in encryption algorithms, it
is by no means the only way to encode or decode information.
One further example is the {\em Tiny Encryption Algorithm} (TEA).  
The C-code for TEA \cite{tea} is publically available and the
algorithm has not been broken -- i.e., there is no known classical
algorithm which finds the key significantly faster 
(with input and output given) than the simple brute-force search.
In its standard form, it encodes 64 input bits using a 128 bit key. 
Encoding is done by an iterative procedure that performs -- depending
on the values of the key bits -- different operations on the input
bits. 
With knowledge of the key, these operations are easily invertible,
whereas without the key the inversion of the encryption becomes
exponentially complex. 
Therefore, extracting the key from a given input and its encrypted
output would open a way to decode all information that has been
encrypted with this key. 

Here we will demonstrate that in principle it is possible -- if input
and output are known (e.g., for a specific communication) -- to encode
the search for the key in the ground state of a quadratic Hamiltonian
of the form (\ref{Equadratic}). 
With the input given by the $n=32$ bit numbers $y^1$ and $z^1$ and the
output by the 32-bit numbers $y^\ell$ and $z^\ell$, the TEA algorithm
establishes a relation between these and the encryption key numbers
$k_a, k_b, k_c, k_d$ by forming the sequence
\bea
\label{Etea}
y^{i+1} = 
\left[S_{4}^-(z^i) + k_a\right] \oplus \left[z^i + s^i\right] \oplus 
\left[S_{5}^+(z^i) + k_b\right],
\nn
z^{i+1} = 
\left[S_{4}^-(y^i) + k_c\right] \oplus \left[y^i + s^i\right]
\oplus\left[S_{5}^+(y^i) + k_d\right],
\nonumber
\eea
where the index $i$ runs from 1 to $\ell-1$.
Here $\oplus$ denotes bitwise addition (XOR), $S^\pm_a$ defines
left- and right- bit-shifting by $a$ digits, and $s^i$ are some 
fixed (known) numbers given by the algorithm.
In its standard form, the algorithm uses $\ell=32$ iterations. 
Obviously, in each of the three different operations used -- 
adding of two numbers, bit shifting and bitwise addition (XOR) -- 
always one of the two operands is defined by the key or by the
encryption algorithm. 
Therefore, knowledge of input, output and the key enables one to
reverse the process (of course, with knowledge of the encryption
algorithm assumed). 
From the previous section it is evident that the addition of two
$n$-bit numbers [none of which is known -- e.g., $S_{4}^-(z^i) + k_a$] 
can be encoded in the ground state of a quadratic Hamiltonian using
just $n$ ancilla bits. 
Using the replacement (\ref{Epenalty_new}), the same is possible for
the XOR operation. 
Note that the bit shifting used does not pose any (theoretical)
problem, since it just changes the index of qubits which have to
interact. 
In each iteration of (\ref{Etea}) one has 4 XOR operations and 4
non-trivial additions (where both addends are unknown). 
Therefore, a quadratic Hamiltonian would not only operate on the key
bits $k_a, k_b, k_c, k_d$ but also on the intermediate results 
$y^i, z^i$ as well as the ancilla qubits. 
In each iteration, $\ord\{n\}$ auxiliary qubits are required, such
that an overall number of $\ord\{n \ell\}$ bits will suffice to encode
the search for the key in the ground state of a quadratic
Hamiltonian. 
{\em Ergo}, similar to factoring, we obtain a quadratic overhead and thus
one might expect that an adiabatic quantum algorithm might find the
key much faster than the Grover (brute-force) search.  



\begin{thebibliography}{9999}

\bibitem{shor}
P.~W.~Shor,
SIAM J.\ Comp.\ {\bf 26}, 1484 (1997).

\bibitem{grover}
L.~K.~Grover,
Phys.\ Rev.\ Lett.\ {\bf 79}, 325 (1997).

\bibitem{nielsen2000}
M.~A.~Nielsen and I.~L.~Chuang,
{\em Quantum Computation and Quantum Information},
Cambridge University Press, Cambridge (2000).

\bibitem{aliferis2006a}
P. Aliferis, D. Gottesman, and J. Preskill,
Quant.\ Inf.\ Comp.\ {\bf 6}, 97-165 (2006).

\bibitem{farhi}
E.~Farhi {\em et al.},
Science {\bf 292}, 472 (2001).

\bibitem{farhi_ax}
E.~Farhi {\em et al.},
e-print: {\tt quant-ph/0001106} (2001).

\bibitem{aharonov}
D.~Aharonov {\em et al.},
45th Annual IEEE Symposium on Foundations of Computer Science, 
42-51 (2004);
e-print: {\tt quant-ph/0405098}.

\bibitem{sarandy2004}
M.~S.~Sarandy, L.-A.~Wu and D.~A.~Lidar,
Quant.\ Inform.\ Proc.\ {\bf 3}, 331 (2004).

\bibitem{childs2001a}
A.~M.~Childs, E.~Farhi, and J.~Preskill,
Phys.\ Rev.\ A {\bf 65}, 012322 (2001).

\bibitem{sarandy2005a}
M.~S.~Sarandy and D.~A.~Lidar,
Phys.~Rev.~A {\bf 71}, 012331 (2005).

\bibitem{sarandy2005b}
M.~S.~Sarandy and D.~A.~Lidar,
Phys.~Rev.~Lett.\ {\bf 95}, 250503 (2005).

\bibitem{roland2005a}
J.~Roland and N.~J.~Cerf,
Phys.\ Rev.\ A {\bf 71}, 032330 (2005).

\bibitem{aberg2005a}
J.~{\AA}berg, D.~Kult, and E.~Sj\"oqvist,
Phys.~Rev.~A {\bf 71}, 060312(R) (2005).

\bibitem{aberg2005b}
J.~{\AA}berg, D.~Kult, and E.~Sj\"oqvist,
Phys.~Rev.~A {\bf 72}, 042317 (2005).

\bibitem{thunstroem2005a}
P.~Thunstr\"om, J.~{\AA}berg, and E.~Sj\"oqvist,
Phys.~Rev.~A {\bf 72}, 022328 (2005).

\bibitem{shenvi2003a}
N.~Shenvi, K.~R.~Brown, and K.~B.~Whaley,
Phys.~Rev.~A {\bf 68}, 052313 (2003).

\bibitem{tiersch}
M.~Tiersch and R.~Sch\"utzhold,
Phys.\ Rev.\ A {\bf 75}, 062313 (2007).

\bibitem{mostame2007a}
S.~Mostame, G.~Schaller, and R.~Sch\"utzhold,
Phys.\ Rev.\ A {\bf 76}, 030304(R) (2007).

\bibitem{schaller2006b}
G.~Schaller, S.~Mostame, and R.~Sch\"utzhold,
Phys.\ Rev.\ A {\bf 73}, 062307 (2006).

\bibitem{jansen2007a}
S. Jansen,  M. B. Ruskai, and R. Seiler,
J.\ Math.\ Phys.\ {\bf 48}, 102111 (2007).

\bibitem{roland2002}
J.~Roland and N.~J.~Cerf,
Phys.\ Rev.\ {\bf 65}, 042308 (2002).

\bibitem{znidaric+horvat}
M.~\u{Z}nidari\u{c} and M.~Horvat, 
Phys.\ Rev.\ A {\bf 73}, 022329 (2006). 

\bibitem{farhi-fail}
E.~Farhi {\em et al.}, 
Int.\ J.\ Quant.\ Inf.\ {\bf 6}, 503-516 (2008).

\bibitem{znidaric}
M.~\u{Z}nidari\u{c},
Phys.\ Rev.\ A {\bf 71}, 062305 (2005).

\bibitem{mosca}
L.~M.~Ioannou and M.~Mosca, 
Int.\ J.\ Quant.\ Inf.\ {\bf 6}, 419 - 426 (2008).

\bibitem{latorre2004a}
J.~I.~Latorre and R.~Orus,
Phys.\ Rev.\ A {\bf 69}, 062302 (2004).

\bibitem{quantum_phase}
R.~Sch\"utzhold and G.~Schaller,
Phys.~Rev.~A {\bf 74}, 060304(R) (2006).

\bibitem{farhi0201031}
E. Farhi, J. Goldstone, and S. Gutmann,
e-print:{\tt quant-ph/0201031} (2002).

\bibitem{farhi0208135}
E. Farhi, J. Goldstone, and S. Gutmann,
e-print:{\tt quant-ph/0208135} (2002).

\bibitem{sachdev}
S.~Sachdev,
{\em Quantum Phase Transitions},
Cambridge University Press, Cambridge (2000).

\bibitem{dziarmaga}
J.~Dziarmaga,
Phys.\ Rev.\ Lett.\ {\bf 95}, 245701 (2005).

\bibitem{schaller2008b}
G. Schaller, 
Phys.\ Rev.\ A {\bf 78}, 032328 (2008).

\bibitem{schuetzhold2008a}
R. Sch\"utzhold, 
J.\ Low Temp.\ Phys.\ {\bf 153}, 228 (2008).

\bibitem{banuls2006a}
M.~C.~Banuls {\em et al.}, 
Phys.\ Rev.\ A {\bf 73}, 022344 (2006). 

\bibitem{young2008a}
A. P. Young, S. Knysh and V. N. Smelyanskiy,
Phys.\ Rev.\ Lett.\ {\bf 101}, 170503 (2008).

\bibitem{childs2002}
A.~M.~Childs {\em et al.}, 
Quant.\ Inf.\ Comp.\ {\bf 2}, 181 (2002).

\bibitem{arpack1998}
R.~B.~Lehoucq, D.~C.~Sorensen, and C.~Yang,
{\em ARPACK Users' Guide: Solution of Large-Scale Eigenvalue Problems
	with Implicitly Restarted Arnoldi Methods},
SIAM (1998); see also 
{\tt http://www.caam.rice.edu/software/ARPACK}.

\bibitem{das2003}
S. Das, R. Kobes, and G. Kunstatter,
J.\ Phys.\ A {\bf 36}, 2839-2845, (2003).

\bibitem{kalapala}
V.~Kalapala and C.~Moore,
e-print: {\tt cs.CC/0508037} (2005).

\bibitem{press1994}
W.~H.~Press {\em et al.}, 
{\em Numerical Recipes in C},
Cambridge University Press, Cambridge (1994).

\bibitem{orus2004a}
R.~Orus and J.~I.~Latorre,
Phys.\ Rev.\  A {\bf 69}, 052308 (2004).

\bibitem{vidal}
G.~Vidal, 
Phys.\ Rev.\ Lett.\ {\bf 91}, 147902 (2003). 

\bibitem{cincio2007a}
L.~Cincio {\em et al.}, 
Phys.\ Rev.\ A {\bf 75}, 052321 (2007).

\bibitem{burges}
C.~J.~C.~Burges,
{\em Factoring as Optimization},
Microsoft Research MSR-TR-2002-83, 
Technical Report (2002).

\bibitem{tea}
D.~J.~Wheeler and R.~M.~Needham,
Lecture Notes in Computer Science {\bf 1008}, 363-366 (1994);
see also {\tt http://www.simonshepherd.supanet.com/tea.htm}.

\end{thebibliography}
\end{document}